\documentclass[%
prd,nofootinbib,showpac,
superscriptaddress,
preprint
]{revtex4-1}
\usepackage{adjustbox}
\usepackage{graphicx}
\usepackage{dcolumn}
\usepackage{bm}

\usepackage[utf8x]{inputenc} 
\usepackage{amsmath,amssymb} 
\usepackage{adjustbox}
\usepackage{color}

\usepackage[colorlinks,citecolor=blue]{hyperref}

\usepackage{float}
\usepackage{rotating}
\usepackage[font=small,labelfont=bf]{caption}
\usepackage{upgreek}
\usepackage{subcaption}

\newcommand{\thetamu}{$\theta_{23}$}
\newcommand{\thetae}{$\theta_{13}$}

\newcommand{\dcp}{$\delta_{\text{CP}}$}

\newcommand{\nova}{NO$\nu$A}
\newcommand\brabar{\raisebox{-4.0pt}{\scalebox{.2}{
\textbf{(}}}\raisebox{-4.0pt}{{\_}}\raisebox{-4.0pt}{\scalebox{.2}{\textbf{)
}}}}

\begin{document}

\preprint{}

\title{Disentangling data contributions to the precision measurement of the largest leptonic mixing angle}

	\author{P. T. Quyen}
	\email{phantoquyen@ifirse.icise.vn}
	\affiliation{\textit{Institute for Interdisciplinary Research in Science and Education,\\ \it{ICISE, Quy Nhon, Vietnam.}}}
	\affiliation{\textit{Graduate University of Science and Technology, Vietnam Academy of Science and Technology, Hanoi, Vietnam.}}
    \author{Son Cao}
	\email{cvson@ifirse.icise.vn}
    \affiliation{\textit{Institute for Interdisciplinary Research in Science and Education,\\ \it{ICISE, Quy Nhon, Vietnam.}}}
	\author{N. T. Hong Van}
	\email{nhvan@iop.vast.vn}
	\affiliation{\textit{Institute of Physics, Vietnam Academy of Science and Technology, Hanoi, Vietnam.}}

\date{\today}
\begin{abstract}%
This study examines the precise measurement of the largest leptonic mixing angle $\theta_{23}$ through the analysis of neutrino oscillation data samples. Our findings indicate that the $\overset{\brabar}{\nu}_{\mu}\rightarrow \overset{\brabar}{\nu}_{e}$ appearance samples, rather than the $\overset{\brabar}{\nu}_{\mu}\rightarrow \overset{\brabar}{\nu}_{\mu}$ disappearance samples, are sensitive to testing the hypothesis of maximal mixing $\theta_{23}=\pi/4$, particularly if $\theta_{23}$ resides in the higher octant and $\sin^2\theta_{23}<0.54$. The former serves as the primary source for determining the octant of the $\theta_{23}$ mixing angle; however, the latter remains relevant for wide-band energy experiment like DUNE if $\theta_{23}$ is significantly deviated from the maximal value. In the joint analysis of T2HK and DUNE with ultimate precision 1\% on $\sin^{2}\theta_{13}$, utilizing only appearance sub-samples can exclude the maximal-mixing and determine the actual octant for about 64\% of the currently allowed range of the $\theta_{23}$ angle. We argue, that despite the presence of parameter degeneracy, the precise measurement of $\theta_{23}$ exhibits minimal dependence on other unknown factors, including the CP-violation phase and neutrino mass ordering.
\end{abstract}


\maketitle

\section{Introduction to neutrino oscillation and leptonic mixing}
Neutrinos were initially proposed by W. Pauli~\citep{Pauli:1930pc} in 1930 to resolve the energy conservation in beta decay and subsequently confirmed in 1956~\citep{2} with detection of reactor neutrinos via inverse beta decay. The Standard Model posits that neutrinos are neutral massless particles with spin of $1/2$, and interact exclusively with matter via the weak interaction, which makes them exceedingly
challenging for observation and detection. The groundbreaking discovery of neutrino oscillations~\citep{PhysRevLett.81.1562,Ahmad:2001an, PhysRevLett.89.011301} revealed that neutrinos would have non-zero masses and experience flavor mixing, phenomena not accounted for by the Standard Model. Neutrinos exhibit transitions between different flavors as they travel a finite distance, a phenomenon referred to as neutrino oscillation. In the context of neutrino oscillations, the three mass eigenstates ($v_1$, $v_2$, $v_3$) are related to the three flavor eigenstates ($v_e$, $v_\mu$, $v_\tau$) through a $3 \times 3$ unitary mixing matrix, known as the Pontecorvo–Maki–Nakagawa–Sakata (PMNS) matrix ~\citep{pontecorvo1968neutrino,maki1962remarks}. This matrix is characterized by three mixing angles $(\theta_{12}, \theta_{13}, \theta_{23})$, one Dirac CP phase $\delta_{CP}$, and two Majorana phases $(\alpha_{1}, \alpha_{2})$ assuming neutrinos are naturally Majorana particles. The first four parameters, together with the neutrino mass-squared splittings represented as $\Delta m_{ij}^2 = m_i^2 - m_j^2$ ($i, j = 1, 2, 3$), manifest themselves in the neutrino oscillation probability which can be measured whereas the Majorana phases are insensitive to measurements of this kind.

Over the past few decades, our understanding of neutrino oscillation parameters~\citep{ParticleDataGroup:2024cfk} has been significantly advanced, as assessed through data obtained from solar, atmospheric, reactor, and accelerator-based neutrino experiments. Global neutrino data has established two scales for the neutrino mass-squared splittings with high precision. The smaller of these two, the solar mass-squared splitting $\Delta m_{21}^2$, is approximately $7.4\times10^{-5}eV^{2}/c^{4}$ with 2.8\% uncertainty. The larger, the atmospheric mass-squared splitting $|\Delta m_{32}^2|$, is about $2.5 \times 10^{-3}\text{eV}^2 / c^4$ with 1.1\% uncertainty. The existing data from accelerator-based experiments, T2K~\citep{T2K:2023smv} and NO$\nu$A~\citep{acero2022improved}, as well as atmospheric neutrino experiment, Super-Kamiokande~\citep{Super-Kamiokande:2023ahc} do not provide conclusive evidence regarding the sign of $\Delta m_{32}^2$. Consequently, the neutrino mass ordering (MO), whether \emph{normal} with $m_3 > m_2 > m_1$ or \emph{inverted} $m_2 > m_1 > m_3$, remains unconfirmed.  Upcoming experiments such as Jiangmen Underground Neutrino Observatory (JUNO)~\citep{djurcic2015juno}, Hyper-Kamiokande (T2HK)~\citep{protocollaboration2018hyperkamiokande} and Deep Underground Neutrino Experiment (DUNE)~\citep{DUNE:2020lwj} are anticipated to determine conclusively the neutrino mass ordering. The three leptonic mixing angles are established to be significantly larger ($\theta_{12}\approx \pi/6,\ \theta_{13}\approx \pi/20,\ \theta_{23}\approx \pi/4$) than the corresponding quark mixing angles ($\theta_{12}^{q}\approx \pi/14,\ \theta_{13}^q\approx \pi/896,\ \theta_{12}^q\approx \pi/76$)~\citep{ParticleDataGroup:2024cfk}. The three parameters are measured precisely with few percent of uncertainty, as indicated in Table~\ref{tab:nuoscpara}. Large mixing in leptons can lead to greater CP violation, as measured by the Jarlskog invariant~\citep{Jarlskog:1985ht}, $J_{CP} = \frac{1}{8}\cos\theta_{13}\sin2\theta_{12}\sin2\theta_{13}\sin2\theta_{23}\sin\delta_{CP}$, in the lepton sector compared to the quark sector. The value of the Dirac-type CP-violation phase $\delta_{\text{CP}}$ remains uncertain, although there are some indications~\citep{T2K:2019bcf} available. If the actual value of $\delta_{\text{CP}}$ is close to the maximal value, i.e., $(|\sin \delta_{\text{CP}}|=1)$, a joint analysis~\citep{Cao:2020ans} of the complete dataset from T2K and NO$\nu$A, enhanced by JUNO's sensitivity to neutrino mass ordering, could yield $4\sigma$ significance for excluding CP conservation. The upcoming T2HK~\citep{protocollaboration2018hyperkamiokande} and DUNE~\citep{DUNE:2020lwj} have great potential to discover the CP violation across range over a wide range of $\delta_{\text{CP}}$ possible values. A proposed experiment, the European Spallation Source neutrino Super Beam (ESSnuSB)~\citep{Alekou:2022emd}, aims to measure CP violation with exceptional precision, facilitating the testing of various flavor models and leptogenesis scenarios.

The proximity of the largest leptonic mixing angle $\theta_{23}$ to the maximal mixing, i.e., $\theta_{23}=\pi/4$, is particularly interesting, as it leads to a similar contribution of the second and third flavors into the third mass eigenstate. This may be merely coincidental~\citep{deGouvea:2012ac}, or it could result from an unidentified flavor symmetry, as discussed in the literature reviews~\citep{Branco:2011zb,Feruglio:2019ybq} and vast references cited herein. The precise measurement of this angle is crucial for testing certain categories of lepton flavor models~\citep{Cao:2024ptn}. This study analytically and numerically investigates the impact of neutrino oscillation data samples on the precision measurement of $\theta_{23}$ angle, with particular emphasis on the vicinity of the maximal mixing $\theta_{23}=\pi/4$.  This sensitivity breakdown study should be applied in practical data analysis, given that combined data analyses typically assume the completeness of the PMNS matrix, which remains inadequately established. 

\begin{table*}
    \centering
    \begin{tabular}{l|c|c|c}
    \hline\hline
    Parameter & Best fit & $3\sigma$ C.L. range &Precision\\\hline
    $\sin^{2}\theta_{12}$ & $0.303$ & 0.270-0.341& $\delta(\theta_{12}) = 2.2\%$\\ 
    $\sin^{2}\theta_{13}(\times 10^{-2})$ & 2.203 & 2.0-2.4 & $\delta(\theta_{13}) = 1.3\%$ \\
    $\sin^{2}\theta_{23}$ & 0.572 & 0.406-0.620 & $\delta(\theta_{23}) = 2.3\%$ \footnote{Precision is calculated using 1$\sigma$ from the best-fit value; however, it is important to note that the data do not statistically exclude the lower octant of $\theta_{23}$.}\\ 
    $\delta_{CP}(^{\circ})$ & 197 & 108-404 & \text{\emph{---}}\\
    $\Delta m^{2}_{21} (10^{-5}\text{eV}^{2}/c^4)$ & 7.41 & 6.82-8.03 & $\delta(\Delta m^{2}_{21}) =  2.8\%$\\
    $\Delta m^{2}_{31} (10^{-3}\text{eV}^{2}/c^{4})$ & 2.511 & 2.428-2.597 & $\delta(\Delta m^{2}_{31}) = \text{1.1}\%$\\\hline\hline
    \end{tabular}
    \caption{Global constraint on oscillation parameters assuming \emph{normal} mass ordering, based on data from NuFit 5.2, as reported in Ref.~\citep{Esteban:2020cvm} using information available in November 2022.}
    \label{tab:nuoscpara}
\end{table*}

This paper is structured as follows.  Section~\ref{sec:samples} presents the existing understanding of the mixing angle $\theta_{23}$, emphasizing methods for measuring and improving the precision of $\theta_{23}$ through the sensitivity analysis of oscillation samples. Section~\ref{sec:exp} outlines the experimental setup and physics potential of accelerator-based neutrino experiments to measure the mixing angle $\theta_{23}$. Significance of physical impact of oscillation data samples on the precision of $\theta_{23}$ measurements are mainly discussed in Sections~\ref{sec:res}. We conclude and highlight the implications of these findings in Section~\ref{sec:cons}. 
\section{Contributions from oscillation data samples to $\theta_{23}$ measurement}\label{sec:samples}

\begin{figure*}
   \begin{minipage}{0.5\textwidth}
     \centering
     \captionsetup{width=.9\linewidth}
     \includegraphics[width=1\linewidth]{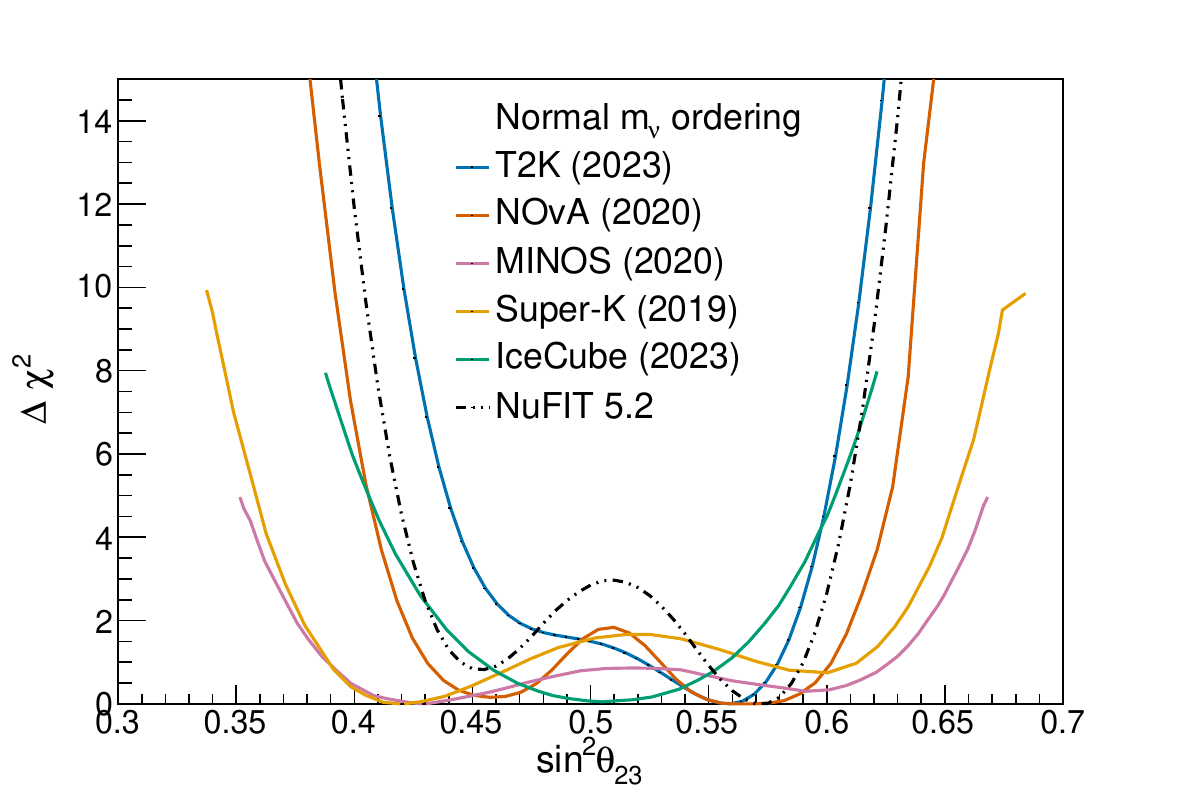}
     \caption{Constraints on mixing angle $\theta_{23}$ from various neutrino experiments, including Super-K~\citep{Super-Kamiokande:2019gzr}, MINOS(+)~\citep{MINOS:2020llm}, NO$\nu$A~\citep{NOvA:2021nfi}, T2K~\citep{T2K:2023smv}, and IceCube~\citep{IceCube:2023ewe}. A global analysis~\citep{Esteban:2020cvm} with NuFIT 5.2 is also presented.}\label{fig:th23global}
   \end{minipage}\hfill
   \begin{minipage}{0.5\textwidth}
     \centering
     \captionsetup{width=.9\linewidth}
     \includegraphics[width=1.\linewidth]{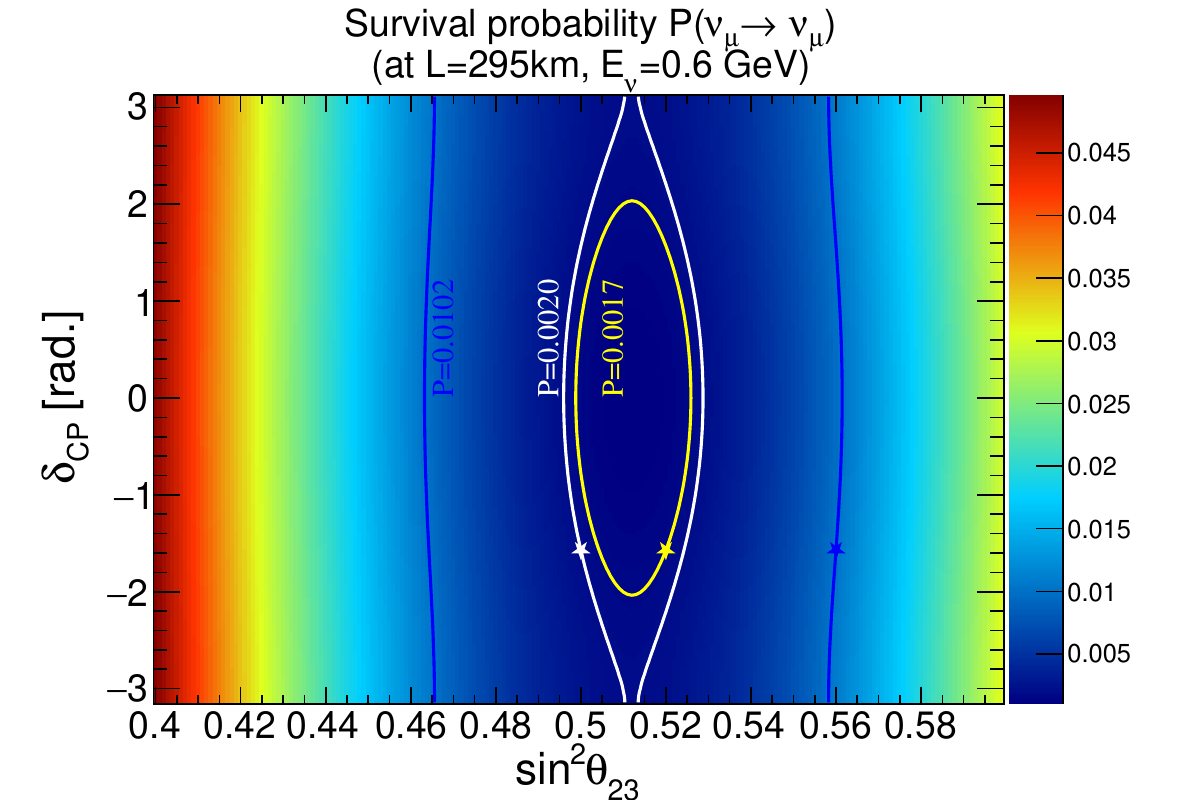}
     \caption{The \emph{disappearance} probability  as a function of $\sin^{2}\theta_{23}$ and $\delta_{CP}$. $\sin^2\theta_{13}=0.02203$ is assumed. The iso-probability curves are calculated at three points of $\sin^2\theta_{23}$=(0.50, 0.52, 0.56) and $\delta_{CP}= -\pi/2$.}\label{fig:prob_numu2numu}
   \end{minipage}
\end{figure*}

Current knowledge of the leptonic mixing angle $\theta_{23}$ is largely derived from atmospheric neutrino experiments and accelerator-based neutrino experiments. Recent measurements of this angle $\theta_{23}$ by various experiments~\citep{Super-Kamiokande:2019gzr,MINOS:2020llm,NOvA:2021nfi,T2K:2023smv,IceCube:2023ewe} and a global data analysis~\citep{Esteban:2020cvm} are presented in Fig.~\ref{fig:th23global}. All experiments show a strong agreement with the maximal mixing hypothesis  $(\theta_{23}=\pi/4)$. A recent joint analysis of T2K and \nova~\citep{t2knovaseminar} shows a mild preference for lower octant of \thetamu. However, when using a reactor-based measurement on \thetae\ as an external constraint, the preference is shifted towards the higher octant. This indicates that measuring \thetamu\ precisely is not straightforward and that it is important to carefully examine the impact of individual data samples, the parameter degeneracy inherent in each measurement, and the assumptions made in a combined data analysis.

\noindent In accelerator-based long-baseline experiments using intense and almost pure muon (anti-muon) neutrino sources, the $\theta_{23}$ value can be measured using two types of oscillation data samples: (i) \emph{appearance} samples  resulting from $\overset{\brabar}{\nu}_{\mu}\rightarrow \overset{\brabar}{\nu}_{e}$ exclusive transition, and (ii) \emph{disappearance} samples collected from $\overset{\brabar}{\nu}_{\mu}\rightarrow \overset{\brabar}{\nu}_{\mu}$ inclusive transition. Considering placing a neutrino detector at the oscillation maximum, i.e., $\Delta_{31}\equiv 1.27\times \Delta m^2_{31} [\text{eV}^2]\frac{L[\text{km}]}{E[\text{GeV}]} \approx \pi/2$, the \emph{disappearance} probabilities~~\citep{suekane2015neutrino} in vacuum are given by: 
\begin{widetext}
\begin{align} \label{mu2muprob}
    P(\overset{\brabar}{\nu}_{\mu}\rightarrow \overset{\brabar}{\nu}_{\mu})|_{(\Delta_{31}\approx \pi/2)} & \approx 1 - \left(\cos^{4}\theta_{13}\sin^{2}2\theta_{23} + \sin^{2}2\theta_{13}\sin^{2}\theta_{23}\right)\sin^{2}\Delta_{31} \nonumber\\
    &+ \epsilon \Delta_{31}\sin2\Delta_{31}\left(\cos^{2}\theta_{12}\sin^{2}2\theta_{23}-\sin^{2}\theta_{23}J_{123}\cos\delta_{CP}\right),
\end{align}
\end{widetext}
\noindent where 
%
$J_{123} = \sin2\theta_{12}\sin2\theta_{13}\sin2\theta_{23}$, and $\epsilon \equiv \frac{\Delta m^{2}_{21}}{\Delta m^{2}_{31}} \sim 0.03$. The leading non-constant term in Eq.~(\ref{mu2muprob}) is driven by $\sin^{2}2\theta_{23}$ since both $\cos^4\theta_{13}$ and $\sin^2\Delta_{31}$ are close to unity, while other terms are suppressed by the smallness of \thetae, $\epsilon$, and $\sin 2\Delta_{31}$. Analytically, this results in a degeneracy in determining the precise value of \thetamu\ given a fixed value of measurable probability $P(\overset{\brabar}{\nu}_{\mu}\rightarrow \overset{\brabar}{\nu}_{\mu})$. The \emph{disappearance} probability for the experimental setup of T2K and T2HK experiments with a baseline of 295~km and neutrino energy of 0.6~GeV is illustrated in Fig.~\ref{fig:prob_numu2numu}. We choose to illustrate on $(\sin^2\theta_{23},\delta_{\text{CP}})$ parameter space since relatively small dependence of \emph{disappearance} probability on these two parameters, leading some non-eligible sensitivity to measure leptonic CP violation~\citep{Denton:2023qmd}. The solid lines in the plot present iso-probability curves, which mean that all points along these lines yield the same probability value.  Fig.~\ref{fig:prob_numu2numu} illustrates that when \thetamu\ deviates notably from $\pi/4$, for example $\sin^2\theta_{23}=0.56$ as presented by blue star, two distinct iso-probability curves show up in both \emph{true} higher octant and \emph{wrong} lower octant. Consequently, it is generally acknowledged that \emph{disappearance} samples are insensitive to the octant of \thetamu. Another striking feature of the \emph{disappearance} probability is the presence of a closed iso-probability curve when \thetamu\ is close to $\pi/4$. This leads to a diminished sensitivity to both two hypothesis tests regarding the exclusion of the maximal mixing and wrong-octant, as confirmed in simulated results presented in Section~\ref{sec:res}. 

\begin{figure*}
\includegraphics[width=0.48\textwidth]{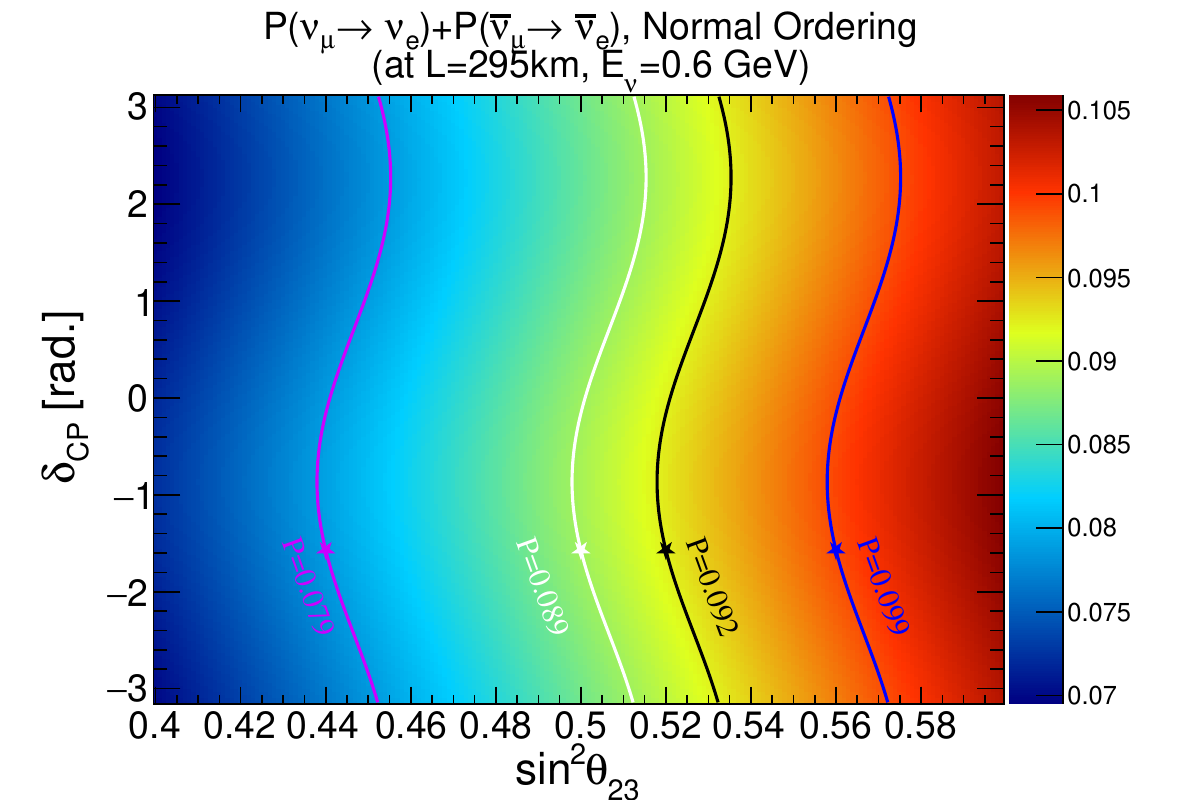}
\includegraphics[width=0.48\textwidth]{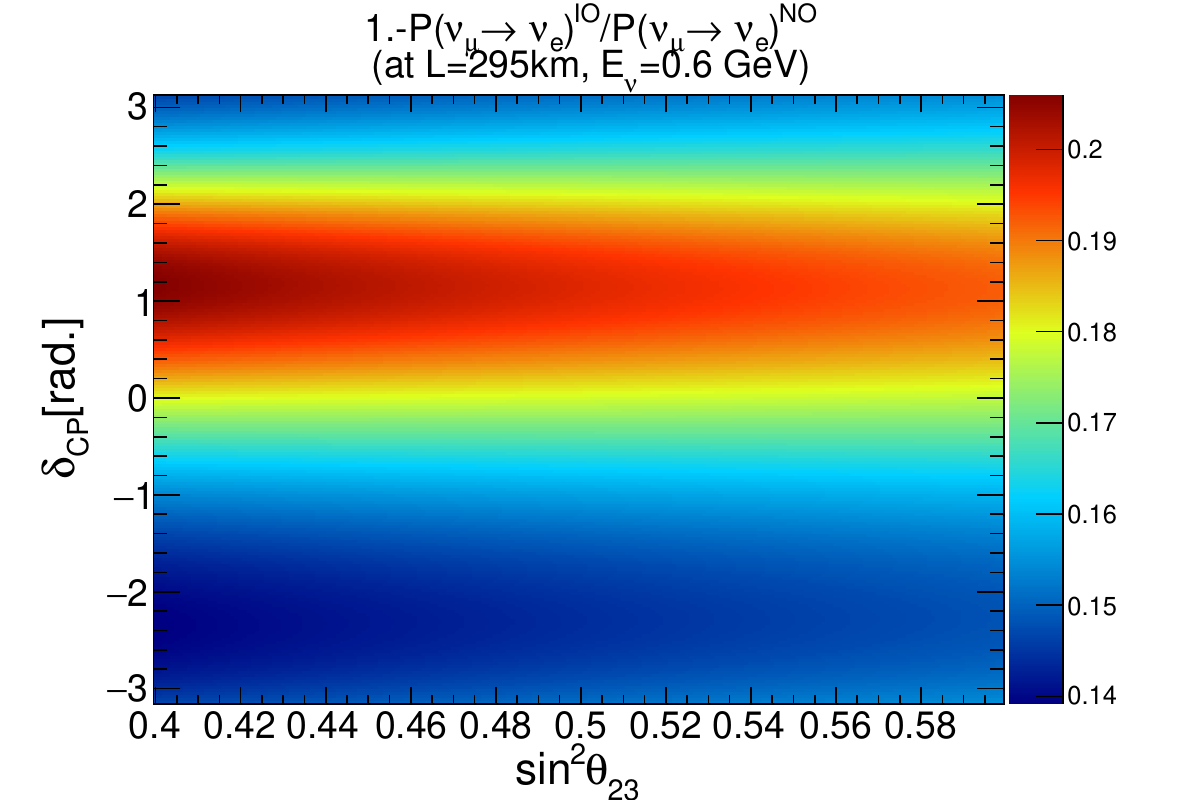}
\caption{\label{fig:prob_numu2nue} The left panel illustrates the sum of \emph{appearance} probabilities for neutrinos and anti-neutrinos as a function of $\sin^{2}\theta_{23}$ and $\delta_{CP}$, under the assumption of \emph{normal} ordering. The iso-probability curves are calculated for four values of $\sin^2\theta_{23}$=(0.44, 0.50, 0.52, 0.56) and $\delta_{CP}= -\pi/2$. The right panel displays the relative difference in \emph{appearance} transitions between \emph{normal} and \emph{inverted} mass orderings.}
\end{figure*}

The \emph{appearance} data samples, in which the statistics are proportional to the  $\overset{\brabar}{\nu}_{\mu}\rightarrow \overset{\brabar}{\nu}_{e}$ transition probabilities, are recognized for its enhanced sensitivity to actual octant of \thetamu. To account for interaction of electron (anti-)neutrinos with matter along the propagation path from the production source to detection location, an effective matter-induced potential $V_{\text{mat.}}(x)= \pm \sqrt{2}G_FN_e$, depending on Fermi coupling constant $G_F$ and electron density $N_e$, is introduced into the Hamiltonian.  With detector placed at the oscillation maximal $\Delta_{31}\approx \pi/2$, the \emph{appearance} probabilities~\citep{Akhmedov:2004ny} can be approximated as 
\begin{widetext}
    \begin{align} \label{eq:probnumutonue}
&P(\overset{\brabar}{\nu}_{\mu}\rightarrow \overset{\brabar}{\nu}_{e})|_{(\Delta_{31}\approx \pi/2)} \approx 4\sin^2\theta_{13}\sin^2\theta_{23}\frac{\sin^2(A-1)\Delta_{31}}{(A-1)^2}\\ \nonumber
&+2\epsilon J_{123}\cos(\Delta_{31}\pm\delta_{\text{CP}})\frac{\sin A\Delta_{31}}{A}\frac{\sin(A-1)\Delta_{31}}{A-1} +\epsilon^2\sin^22\theta_{12}\cos^2\theta_{23}\frac{\sin^2 A\Delta_{31}}{A^2}
\end{align}
\end{widetext}

\noindent where $A\equiv V_{mat.}.L/2\Delta_{31}$, and the $+(-)$ sign is taken for $\nu (\bar{\nu})$ oscillation respectively. The first and also the leading term in Eq.~(\ref{eq:probnumutonue}) is proportional to $\sin^{2}\theta_{23}\sin^{2}\theta_{13}$ and thus, these appearance samples are sensitive to the genuine $\theta_{23}$ octant, assuming the \thetae\ known with high precision. Neutrino experiments that measure the \emph{appearance} probabilities are enable to determine \thetae\ with high precision and must rely on reactor-based measurements of $\overline{\nu}_{e}\rightarrow \overline{\nu}_e$ sample. Due to the degeneracy between \thetamu\ and \thetae, as discussed in Appendix~\ref{sec:AppDeg}, it is essential to verify the consistency of \thetae\ measurements from  accelerator-based and reactor-based experiments before integrating them in a unified framework. Moreover, the \emph{appearance} probability in Eq.~\ref{eq:probnumutonue} exhibits a strong dependence on other unknown factors, including the value of CP violation phase \dcp\ and neutrino mass ordering, as discussed in Appendix~\ref{sec:AppDeg}. Simultaneously extracting multiple parameters from measurable \emph{appearance} probabilities may yield multiple spurious solutions. In the context of precision measurement of \thetamu, it is noteworthy that the sum of the \emph{appearance} probabilities from neutrino and anti-neutrino transitions, as presented in the left plot of Fig.~\ref{fig:prob_numu2nue}, exhibits marginal dependence on the actual value of \dcp. This suggests, and will be validated through numerical simulation in Section~\ref{sec:res}, that in presence of both neutrino and anti-neutrino \emph{appearance} data samples, the precise measurement of \thetamu\ is unaffected by unknown values of \dcp. 

\noindent The effect of neutrino mass ordering is also presented in the \emph{appearance} probability due to the presence of the matter effect. The profiles of $P(\overset{\brabar}{\nu}_{\mu}\rightarrow \overset{\brabar}{\nu}_{e})$ for two mass ordering scenarios are shown in Appendix~\ref{sec:AppDeg}. The right plot of Fig.~\ref{fig:prob_numu2nue} shows that the relative difference between the appearance transitions in 
the normal and inverted ordering exhibits small variation as a function of \thetamu\ although it depends on the true value of \dcp. This, further supported by numerical simulation in Section~\ref{sec:res}, indicates that the precision measurement of \thetamu\ is not influenced by the true neutrino mass ordering. 

\section{Accelerator-based neutrino experiments targeting $\theta_{23}$ precision and beyond}\label{sec:exp}
 A general approach to measuring neutrino oscillation and extracting oscillation parameters including $\theta_{23}$ mixing angle is to compare the initial neutrino flavor composition with the flavor composition after some neutrino propagation distance. We consider a flux of $\alpha$-flavor neutrino ($\Phi_{flux}^{\nu_{\alpha}}$), defined as a function of the \emph{true} neutrino energy $(E_{\nu}^{true})$, prepared for experiment. A neutrino detector is constructed to observe $\beta$-flavor neutrino at a distance $L$, so the neutrinos experience an oscillation probability $P(\nu_{\alpha}\rightarrow \nu_{\beta}|E_{\nu}^{true},\vec{o})$ where $ \vec{o} = (\Delta m^2_{21}, \Delta m^2_{31};\ \theta_{12},\theta_{13},\theta_{23},\delta_{CP} )$ represents the vector of oscillation parameters. The detector is essentially characterized by the mass $M_{det.}$, detection efficiency of identifying $\beta$-flavor neutrino $\epsilon_{det.}^{\nu_{\beta}} (E_{\nu}^{reco.})$ as a function of reconstructed-energy $(E_{\nu}^{reco.})$, and true-to-reconstructed energy smearing response $S_{det.}$. The predicted number of $\nu_{\beta}$ events in $i^{th}$ bin of reconstructed neutrino energy for an exposure time $T$, depending on the interaction cross section $\sigma_{int.}^{\nu_{\beta}}$, can be computed as
\begin{widetext}
     \begin{align}
N_{i,\ pred.}^{\nu_{\beta}}(E_{\nu}^{reco.},\vec{o}) = & T\cdot \int_{E_{\nu}^{reco}-\delta E}^{E_{\nu}^{reco}+\delta E} dE_{\nu}^{reco} \int dE_{\nu}^{true} \cdot \Phi_{flux}^{\nu_{\alpha}}(E_{\nu}^{true})\cdot P(\nu_{\alpha}\rightarrow \nu_{\beta}|E_{\nu}^{true},\vec{o})\\ \nonumber
&\cdot \sigma_{int.}^{\nu_{\beta}}(E_{\nu}^{true})\cdot M_{det.}\cdot S_{det.}(E_{\nu}^{true.}, E_{\nu}^{reco.}) \cdot \epsilon_{det.}^{\nu_{\beta}} (E_{\nu}^{reco.}) 
\end{align}
 \end{widetext}

 \noindent Neutrino experiments measure spectrum $N_{i,\ data}^{\nu_{\beta}}(E_{\nu}^{reco.})$ and compare $N_{i,\ pred.}^{\nu_{\beta}}(E_{\nu}^{reco.},\vec{o})$ to extract vector of oscillation parameters $\vec{o}$. The analysis is complex, and often a numerical simulation is employed to account for many sources of uncertainty in comprehending the neutrino flux, neutrino interactions, and detector responses.

 As shown in Fig.~\ref{fig:th23global}, two on-going accelerator-based long-baseline experiments T2K and NO$\nu$A are among the key players to measure precisely the leptonic mixing angle \thetamu. A joint analysis~\citep{Cao:2020ans} of simulated \emph{disappearance} and \emph{appearance} datasets of T2K and \nova\ is anticipated to conclude the $\theta_{23}$ octant up to a 3$\sigma$ C.L. if $\sin^{2}\theta_{23}\geq 0.56$ or $\sin^{2}\theta_{23}\leq 0.46$.  The upcoming T2HK and DUNE experiments are expected to yield more precise measurements of \thetamu,  while also elucidating the other unknowns such as CP violation and neutrino mass ordering. The ESSnuSB proposal, aimed at achieving unprecedented precision in \dcp\ measurement can also provide valuable data for resolving ambiguity in measuring the \thetamu\ mixing angle. In this study, we focus on the physics potentials of T2HK, DUNE, and ESSnuSB, which possess high statistics in both \emph{disappearance} and \emph{appearance} samples, thus offering a unique opportunity for precise measurement of this parameter with clean data samples and consistency verification within a single experiment.  
T2HK~\citep{pontecorvo1968neutrino}, the successor to T2K, represents the third generation of Water Cherenkov detectors and is currently being developed as a leading worldwide experiment in Japan. T2HK is planned to begin data collection in 2027, following the completion of the T2K runs. T2HK will utilize the existing infrastructure used by T2K, including the beam line generated at the J-PARC accelerator facility and near detectors, with hardware upgrades. To produce a more intense neutrino beam, J-PARC accelerator power will be upgraded from the current 800 kW to 1.3 MW, to generate a more intense neutrino beam, yielding data equivalent to $3.2 \times 10^{14}$ protons per pulse by 2026. The far detector is designed to possess a fiducial volume of 187~kton of water in the Cherenkov detector, which is eight times larger than the Super-Kamiokande detector. The far detector will be located 295 km from J-PARC. T2HK could greatly improve the study of leptonic CP violation, mass ordering, and $\theta_{23}$ due to its extensive statistics. Over a decade of operation, T2HK is anticipated to discover CP violation at around 5$\sigma$ C.L. for 57\% of possible values of $\delta_{CP}$, achieving a precision of 3.4\% at $\sin^{2}\theta_{23} = 0.5$. The determination of the $\theta_{23}$ octant and mass ordering can be achieved with a significance more than $3\sigma$ C.L.

DUNE~\citep{DUNE:2020lwj} is hosted by Fermi National Accelerator Laboratory (Fermilab) in Batavia, Illinois, USA. DUNE consists of two detectors: the near detector is situated at Fermilab and the far detector, which has a 40 kton fiducial mass of liquid argon, located at Sanford Underground Research Laboratory in Lead, South Dakota, 1285.9 kilometers from the beam production target. The data corresponding to $7.5\times10^{13}$ protons per pulse can be collected at the 1.2~MW beam power of energetic 120~GeV proton. DUNE is planned to start collecting data in 2031. In 13 years of running, DUNE can discover the CP violation at $3\sigma$ for 75\% of $\delta_{CP}$ values and also provide a great physics potential to definitively determine mass ordering and the precision on oscillation parameters.

ESSnuSB~\citep{Alekou:2022emd} is a prospective superbeam experiment to precisely measure leptonic CP violation. ESSnuSB uses accelerator-based neutrino source from ESS facility. To produce the high-intensity neutrino beam, the ESS proton LINAC plans to achieve the dedicated delivery of a 5~MW intensity beam to have the beam neutrinos with the maximum intensity at 400~MeV. The near detector is at a distance of 250~m from the production point. ESSnuSB uses an identical large water Cherenkov detectors with 538~kton fiducial mass at the far-site of a baseline of $L$ from the neutrino source. At a fixed neutrino energy of 0.24 GeV, the second oscillation maxima corresponding to $L = 360$~km allows ESSnuSB to measure the leptonic CP violation with unprecedented precision about 2.7~times larger than in the first maxima and to reduce the systematic errors. The running time of 5 years in $\nu$-mode and 5 years in $\bar{\nu}$-mode, ESSnuSB can exclude the conserved CP values ($\delta_{CP} = 0, \pm \pi)$ with $12\sigma$ C.L. or greater at the present systematic errors 5\%. Moreover, apart from the CP violation measurement, ESSnuSB will have an ability to discriminate mass ordering with a significance of $5\sigma$ C.L. or greater.  

\begin{table*}
\centering
    \caption{\label{tab:t2hkdune} Main specification of T2HK, DUNE  and ESSnuSB}
   \begin{tabular}{l|c|c|c}    Characteristics & T2HK~\citep{protocollaboration2018hyperkamiokande}   & DUNE~\citep{DUNE:2020lwj} & ESSnuSB~\citep{Alekou:2022emd}\\ \hline
    Baseline & 295~km & 1284.9~km  &360~km\\ 
    Detector type & Water Cherenkov & Liquid Argon TPC  & Water Cherenkov\\ 
    Detector fiducial mass& 187~kton & 40~kton & 538~kton\\
    Beam power & 1.3~MW & 1.2~MW & 5~MW\\ 
    Running time $\nu:\overline{\nu}$& 2.5 yr : 7.5 yr & 6.5 yr : 6.5 yr & 5 yr : 5 yr\\
    \emph{Appearance } data ($\nu_{e} + \bar{\nu_{e}}$)& (2145 + 1144) & (2599 + 479) & (5154 + 291)\\
    
      \emph{Disappearance} data ($\nu_{\mu} + \bar{\nu_{\mu}}$) & (10,330 + 13,707) & (20,169 + 11,779) & (37,762 + 8040) \\
    \hline
  \end{tabular}
\end{table*}

\section{Numerical simulation results for $\theta_{23}$ precision contributions }\label{sec:res}
The GLoBES simulation package~\citep{globes} is subtilized for evaluating numerically the experimental sensitivity to measurements of oscillation parameters, particularly the leptonic mixing angle \thetamu. The setups specified in Ref. \citep{Cao:2024ptn} are employed for the simulations of the T2HK, DUNE, and ESSnuSB experiments. Unless otherwise indicated, the global fit data of NuFIT 5.2, which is compiled in Table ~\ref{tab:nuoscpara}, is assumed to correspond with the true values of the oscillation parameters. In this work, two statistic tests at given oscillation parameter sets specified $\theta_{23}^{\text{True}}$ are performed on the neutrino energy spectra of data samples to assess the accuracy of the $\theta_{23}$ measurement. These tests include (i) the maximal-mixing ($\sin^2\theta_{23}=0.5$) exclusion characterized by $\Delta \chi_1^2$, and (ii) the wrong-octant exclusion, characterized by $ \Delta \chi_2^2$, as formulated by
\begin{widetext}
    \begin{align*}
     \Delta \chi_{1}^{2}|_{\sin^2\theta_{23}^{\text{True}}} & = \chi_{\text{min.}}^{2} (\sin^{2}\theta_{23} = 0.5| \sin^2\theta_{23}^{\text{True}})-\chi^{2}_{\text{min.}}(\sin^{2}\theta_{23}^{\text{Best-fit}}|\sin^2\theta_{23}^{\text{True}}),\\
     \Delta \chi_{2}^{2}|_{\sin^2\theta_{23}^{\text{True}}}& = \chi^{2}_{\text{min.}} (\sin^{2}\theta_{23}^{\text{wrong-octant}}| \sin^2\theta_{23}^{\text{True}})-\chi^{2}_{\text{min.}}(\sin^{2}\theta_{23} ^{\text{Best-fit}}| \sin^2\theta_{23}^{\text{True}}),
\end{align*}
\end{widetext}

\noindent where $\chi^2_{\text{min.}}$ is obtained after a computational minimization over other oscillation parameters. Due to the \thetamu-octant degeneracy as discussed in Section~\ref{sec:samples}, an appropriate range of \thetamu\ values must be considered for this minimization. Unless otherwise specified, a recent $3\sigma$ boundary [0.4, 0.62] is used for our sensitivity estimation. External constraints on \thetae, such as those from reactor-based neutrino experiments, are anticipated to significantly affect measurements utilizing the \emph{appearance} data samples, owing to the first-order dependence of the $\nu_{\mu}\rightarrow \nu_e$ probability on $\sin^2\theta_{13}$. Unless stated otherwise, a 2.6\% uncertainty on $\sin^2\theta_{13}$ is treated as an external constraint in our study.

 \begin{figure*}
\includegraphics[width=0.495\textwidth]{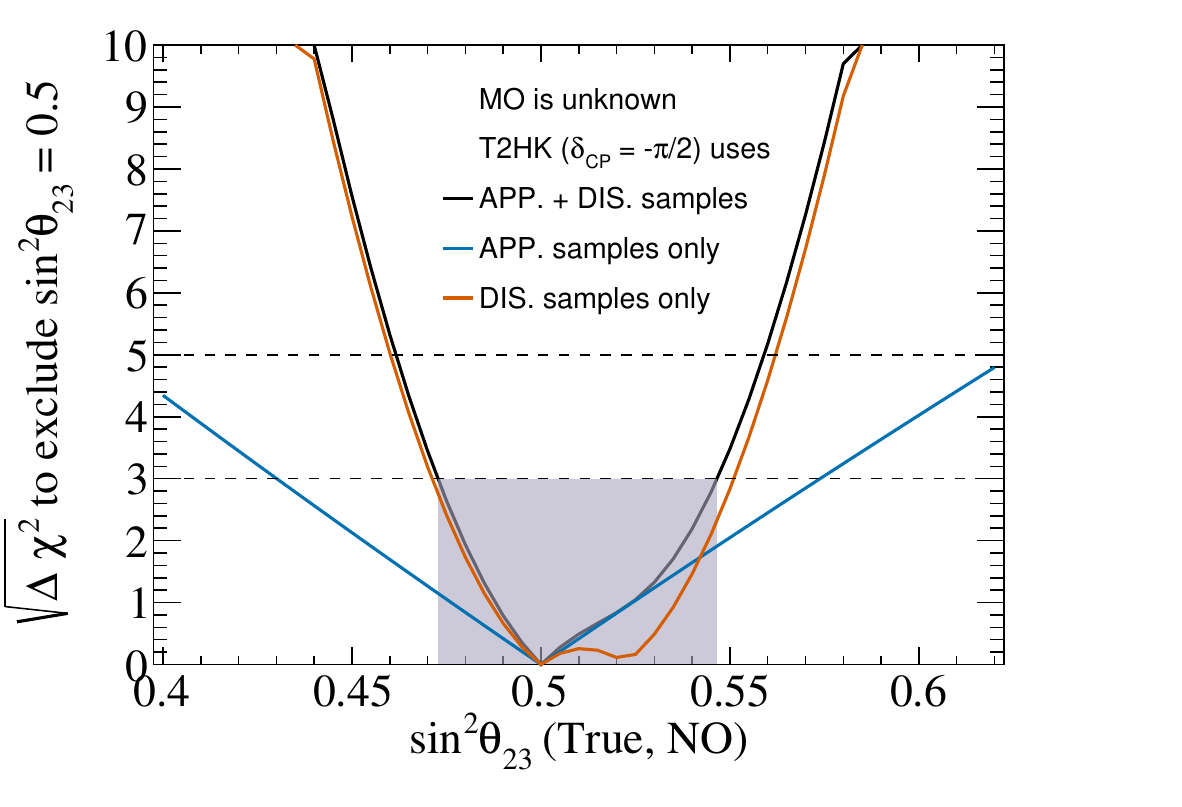}
\includegraphics[width=0.495\textwidth]{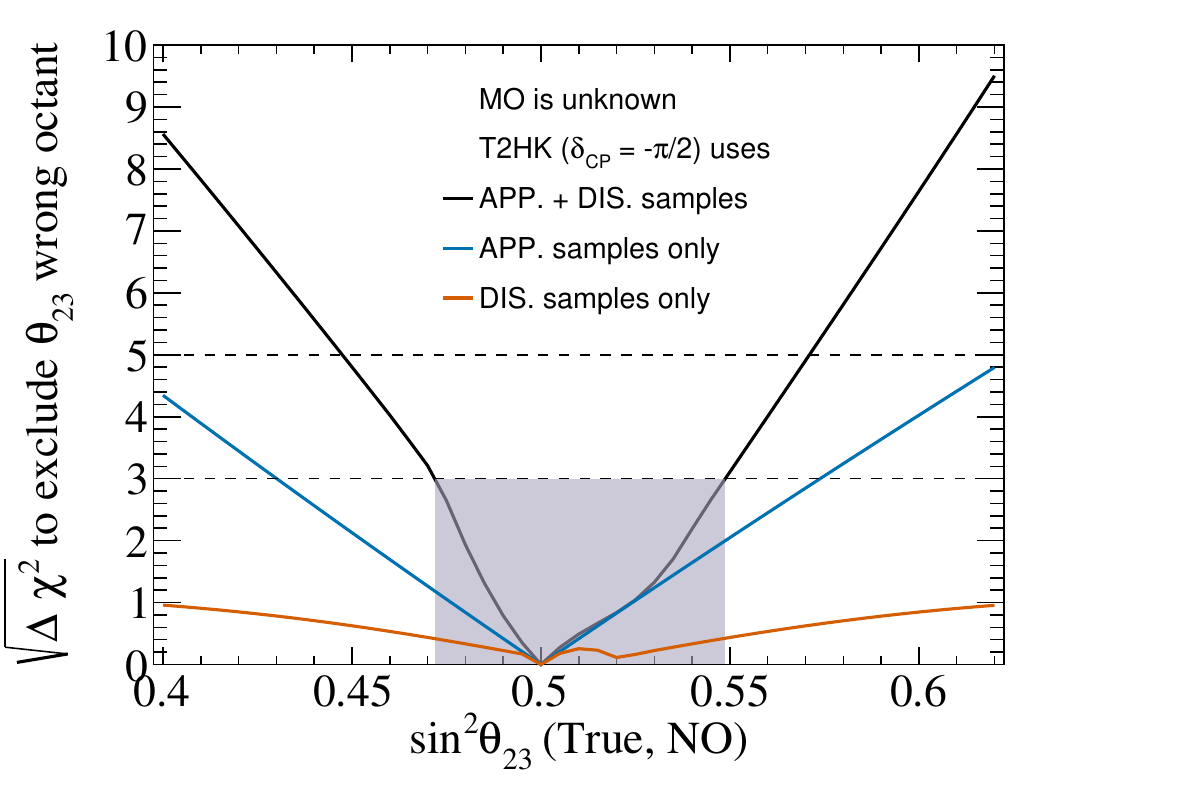}
\includegraphics[width=0.495\textwidth]{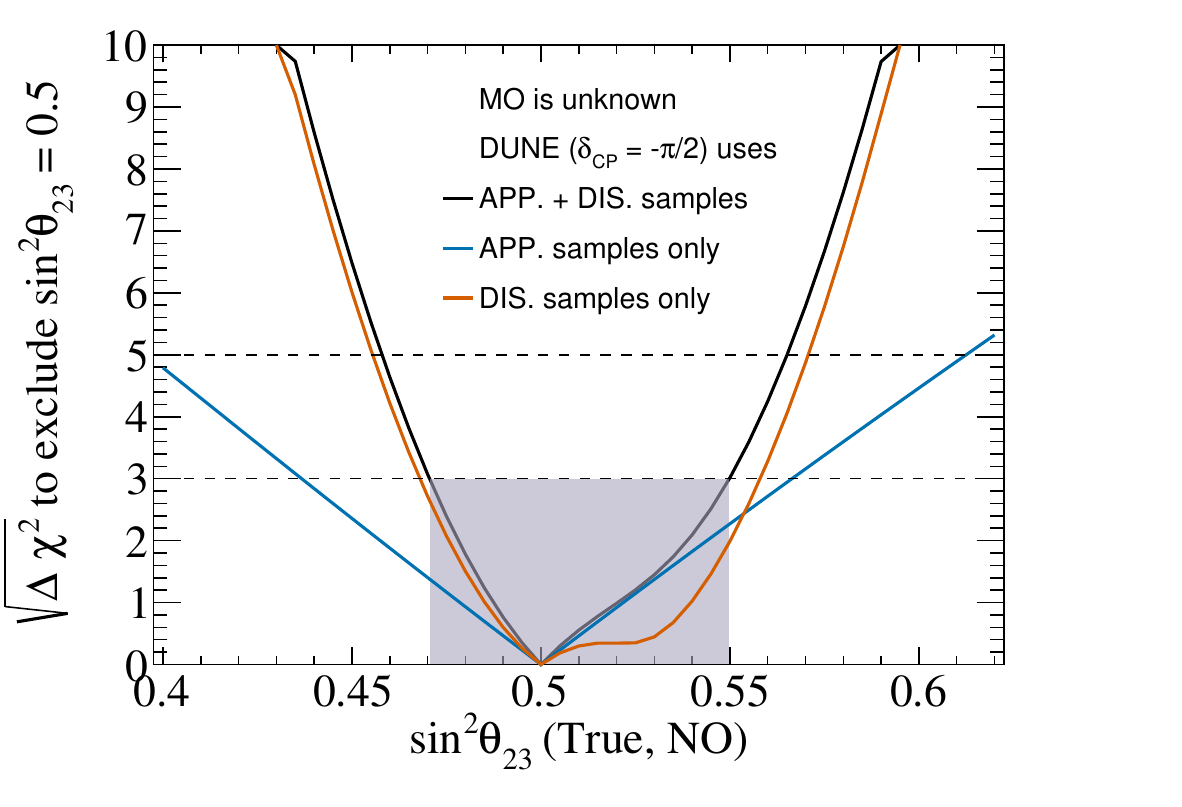}
\includegraphics[width=0.495\textwidth]{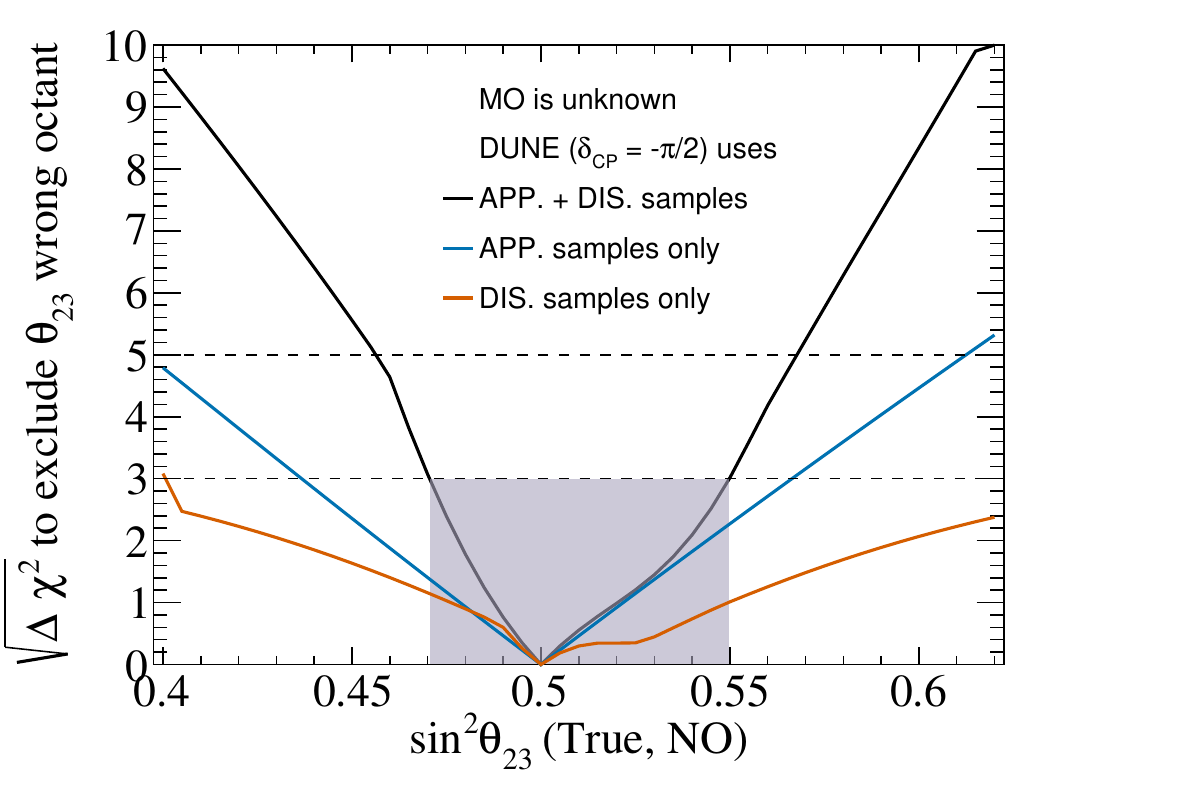}
\includegraphics[width=0.495\textwidth]{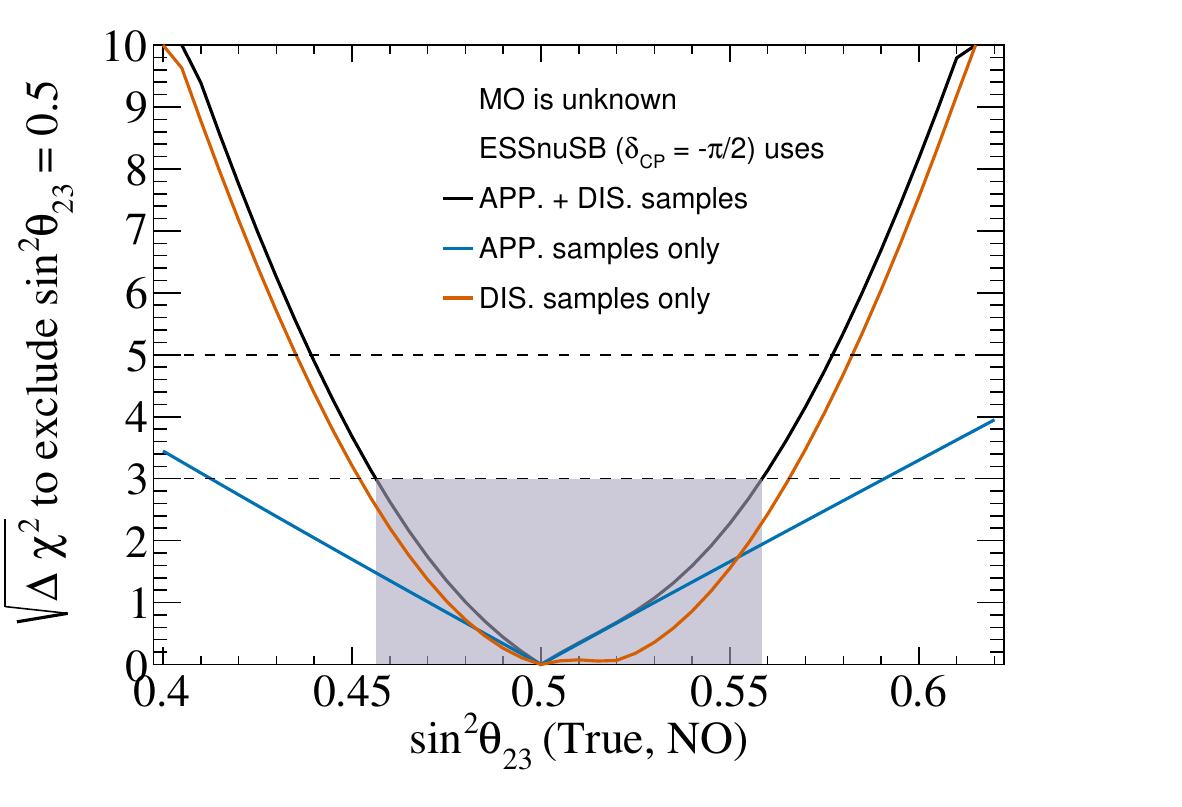}
\includegraphics[width=0.495\textwidth]{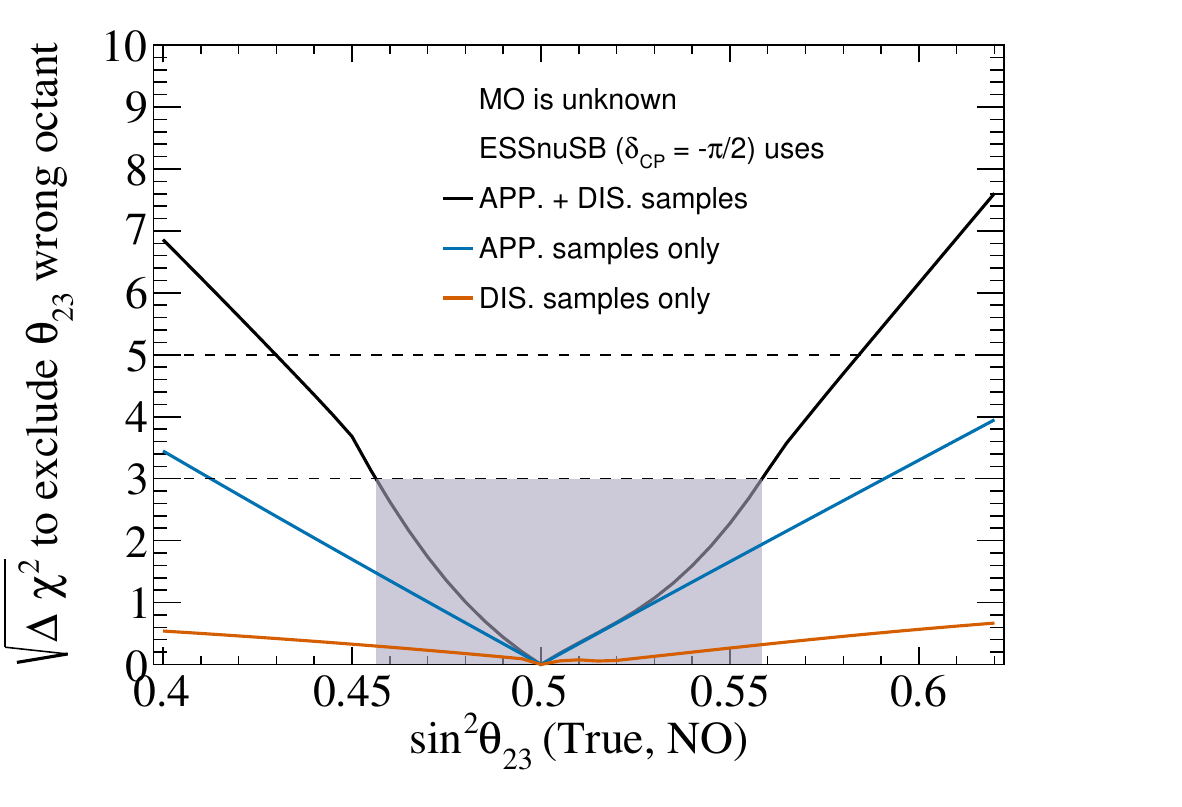}
\caption{\label{fig:octantresolvhyperk} 
The T2HK, DUNE and ESSnuSB's statistical significance to exclude the maximal mixing (left) and wrong-octant (right) as functions of $\sin^{2}\theta_{23}$. Here $\delta_{CP} = -\pi/2$ is set, MO is presumably unknown, and other relevant parameters and their uncertainties are taken from Table~\ref{tab:nuoscpara}. The grey boxes present the maximal-enclosed regions (left) and the octant-blinded region (right) where we could not determine the precise value of $\sin^{2}\theta_{23}$.} 
\end{figure*}

\begin{figure*}
\includegraphics[width=0.495\textwidth]{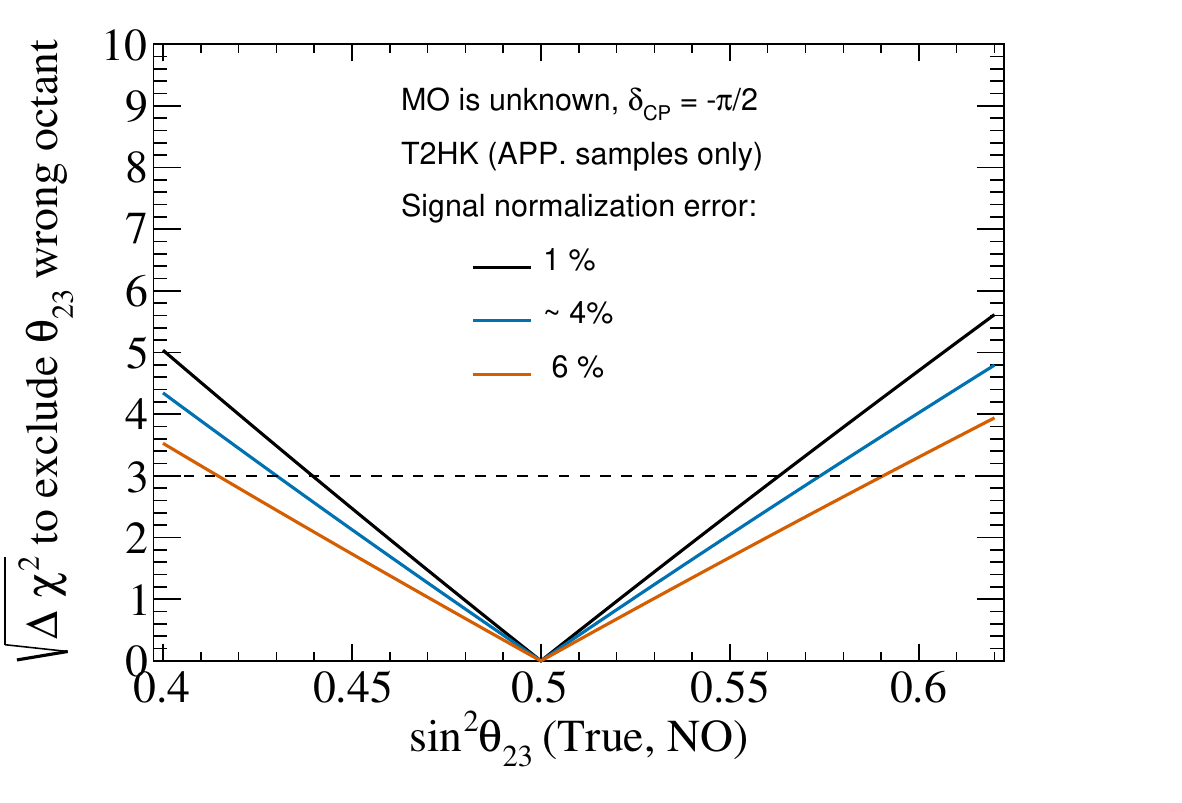}
\includegraphics[width=0.495\textwidth]{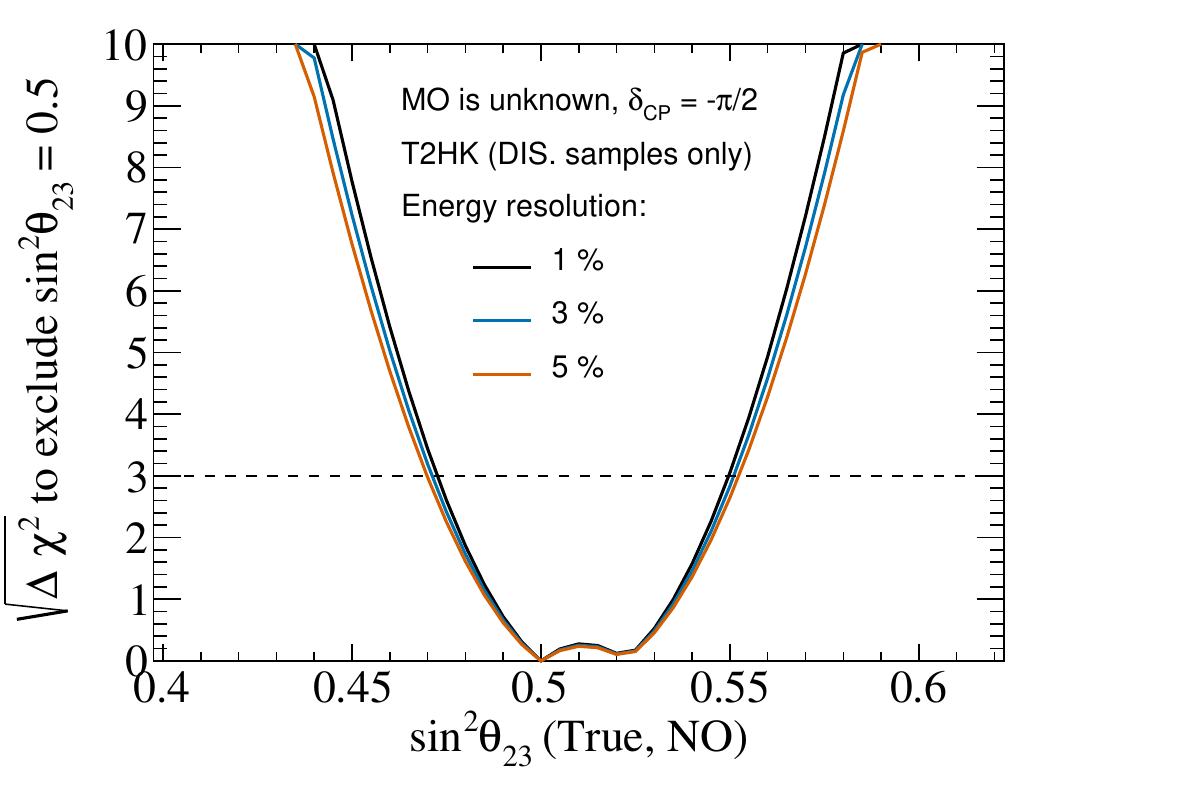}
\caption{\label{fig:syst}The left illustrates the predominant impact of the signal normalization uncertainty on the sensitivity for the wrong-octant exclusion using the \emph{appearance} samples. The right depicts the impact of energy resolution on the maximal-mixing exclusion using the \emph{disappearance} samples. Both are calculated using the T2HK experiment, assuming the mass ordering (MO) is unknown and $\delta_{CP}=-\pi/2$.}
\end{figure*}

\noindent The statistical significance, defined as $\sqrt{\Delta \chi^2}$ for excluding the maximal-mixing and wrong-octant hypotheses in the T2HK, DUNE, and ESSnuSB experiments are presented in Fig ~\ref{fig:octantresolvhyperk}. Compared to T2HK and DUNE, ESSnuSB exhibits a slightly smaller sensitivity for resolving the $\theta_{23}$ ambiguity. T2HK and DUNE can eliminate both the maximal-mixing and wrong-octant hypotheses at a $3\sigma$ C.L. if the actual value of $\sin^2 \theta_{23}$ lies outside the interval [0.47, 0.55], that includes approximately 63\% of the currently allowed range of $\sin^2 \theta_{23}$. The sensitivity of ESSnuSB is limited to the restricted interval of $\sin^2 \theta_{23}$ between [0.46, 0.56], encompassing 45\% of potential values of $\sin^2 \theta_{23}$. The breakdown sensitivities into the contributions from \emph{disappearance} (DIS) and \emph{appearance} (APP) data samples show similar patterns among these three experiments. In T2HK, the contribution from \emph{disappearance} samples for excluding the maximal-mixing hypothesis dominates except in the region where $0.50 \leq \sin^2 \theta_{23} \leq 0.54$, which is primarily driven by \emph{appearance} samples. For excluding the wrong-octant, the contribution of the \emph{appearance} samples becomes more prominent, while \emph{disappearance} samples contribute less in the higher octant for $\sin^2 \theta_{23}$ within the range of (0.50, 0.54). This is due to the octant degeneracy inherent in the \emph{disappearance} probabilities, as analyzed in Section~\ref{sec:samples}. In DUNE, \emph{disappearance} samples predominate in ruling out the maximal-mixing hypothesis, except in the range $0.50 \leq \sin^2 \theta_{23} \leq 0.55$. The sensitivity to the wrong-octant hypothesis is also influenced by \emph{appearance} samples. It is noteworthy in Fig.~\ref{fig:octantresolvhyperk} that the contribution of the disappearance data samples to octant resolution in DUNE is substantially more than that of T2HK and ESSnuSB, primarily due to its broad-band neutrino energy beam. ESSnuSB can ascertain the octant of $\sin^2 \theta_{23}$ if its actual values are out the range of [0.46, 0.56]. The contributions from both \emph{appearance} and \emph{disappearance} samples in ESSnuSB are similar to those in T2HK and DUNE, and align with the oscillation probability analysis presented in Section~\ref{sec:samples}. The impact of systematic errors, such as the signal normalization error, energy resolution, and energy calibration on the \thetamu-octant and maximal-mixing sensitivity has been examined separately for each data sample contribution. The driven systematic is identified as the signal normalization error for the \emph{appearance} contribution, owning to the fact that the deviation of appearance probability in the case of non-maximal value of \thetamu\ from the maximal case has minimal energy dependence in the experimental energy range. The sensitivity of the \emph{disappearance} samples to the maximal-mixing exclusion is slightly affected by the energy resolution near the oscillation dip, as illustrated in Fig.~\ref{fig:syst}. 
   
\begin{figure*}
    \centering
    \includegraphics[width=0.485\linewidth]{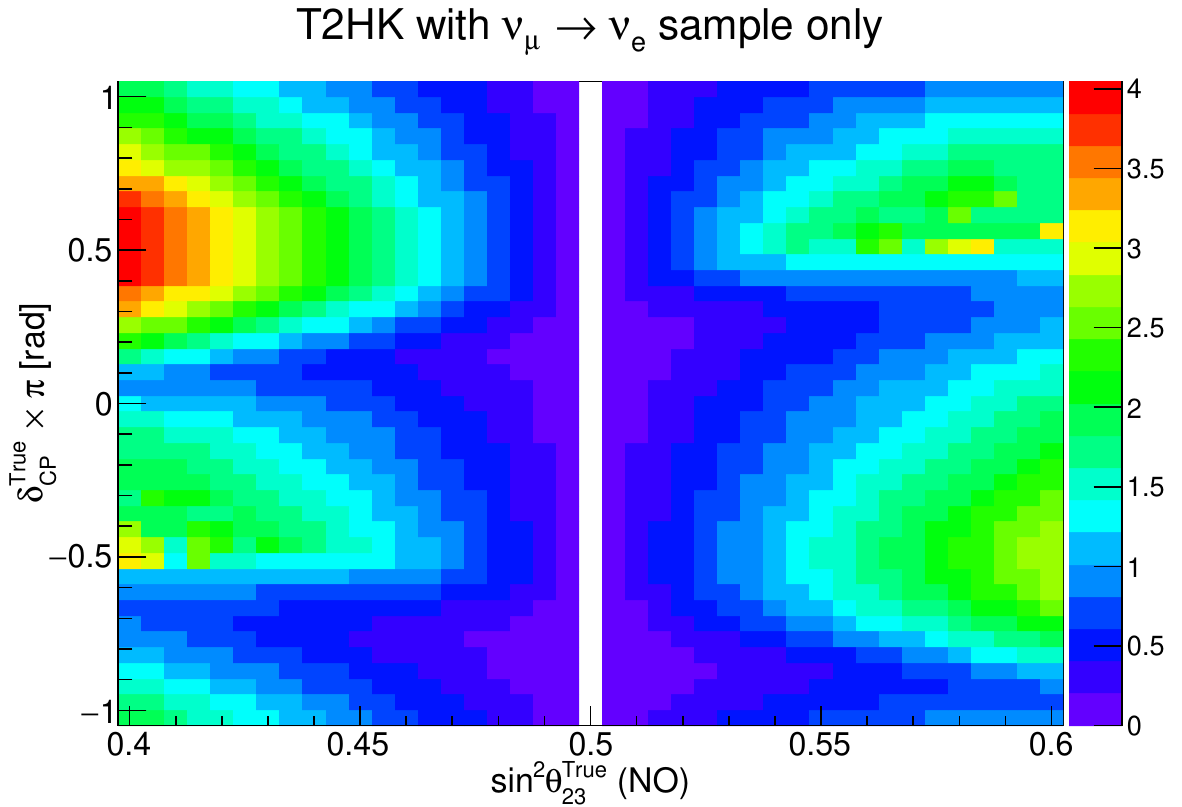}
    \includegraphics[width=0.485\linewidth]{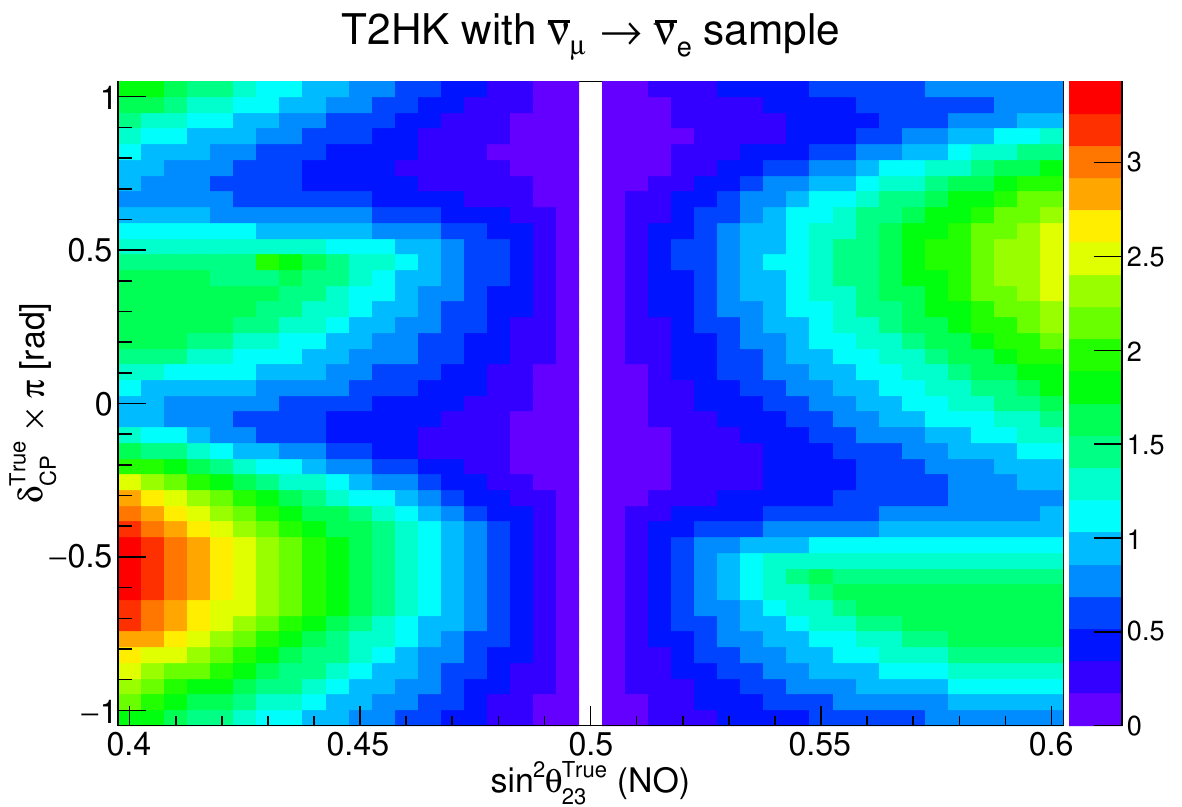}
    \caption{Profiles of statistical significance $\sqrt{\Delta \chi^2}$ for excluding the wrong-octant hypothesis, numerically estimated as a function of ($\sin^{2}\theta_{23}^{\text{True}} - \delta_{CP}^{\text{True}}$). The left and right plots correspond to the $\nu_{e}$ and $\bar{\nu_{e}}$ \emph{appearance} sub-samples of T2HK experiment respectively.}
    \label{fig:chimapnuanti}
\end{figure*}

\begin{figure*}
    \centering
    \includegraphics[width=0.6\linewidth]{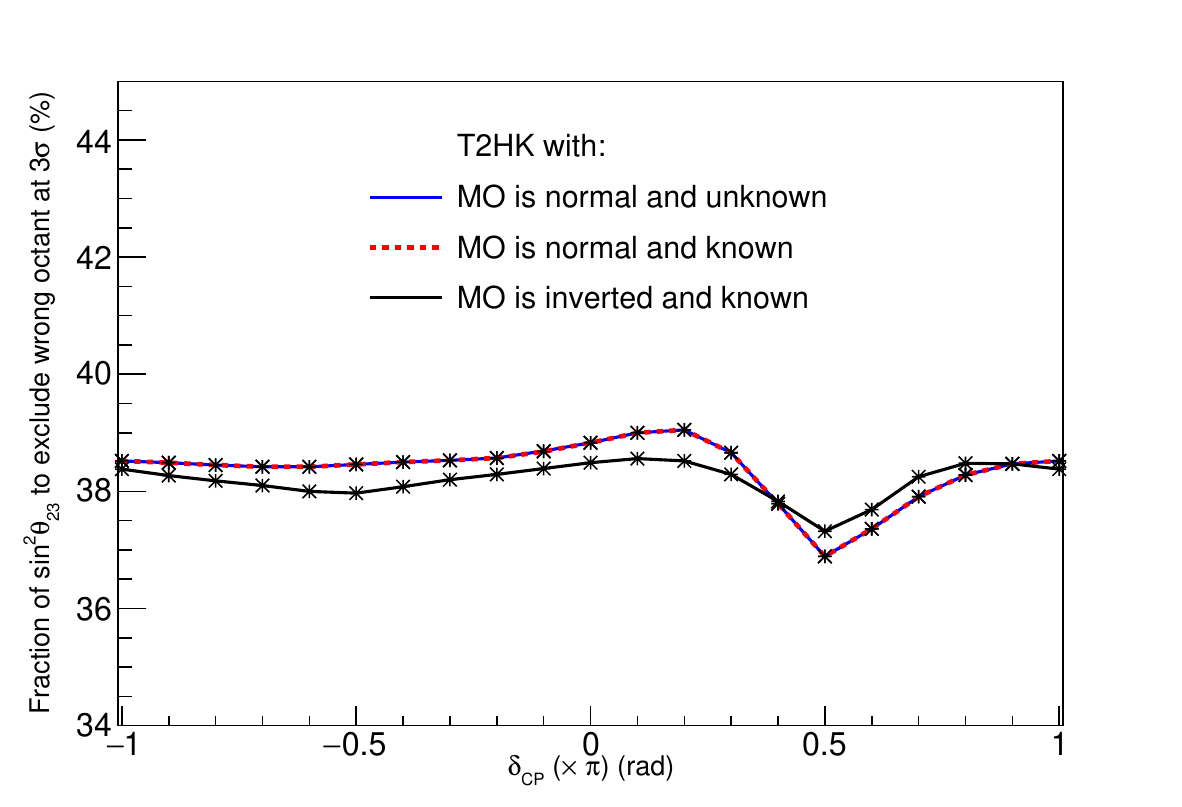}
    \caption{Numerical sensitivity, expressed as a fraction of  $\sin^2\theta_{23}$, for excluding the wrong octant at 3$\sigma$ C.L., shown as a function of true \dcp\ values across three scenarios. This analysis demonstrates that precision measurements of \thetamu\ are largely unaffected by the unknown CP violation phase and mass ordering.  }
    \label{fig:ORvsdcp}
\end{figure*}

The contributions of $\nu$ and $\bar{\nu}$ sub-samples in \emph{appearance} datasets are further analyzed to assess their impact on testing the maximal-mixing and wrong-octant hypotheses, as well as the affect of actual values of the CP-violation phase \dcp. The profiles of statistical significance $\sqrt{\Delta \chi^2}$ presented in Fig.~\ref{fig:chimapnuanti} indicate that the individual sensitivity of $\nu$ and $\bar{\nu}$ sub-samples to exclude the wrong-octant hypothesis largely depends on the actual values of \dcp.  Specifically,  $\nu_{e}$ \textit{appearance} sub-sample, yields higher sensitive to $\theta_{23}$ octant determination in region ($\delta_{CP} \approx \pi/2$, $\sin^{2}\theta_{23} < 0.45$) and ($\delta_{CP} = -\pi/2, \sin^{2}\theta_{23} > 0.55)$. On the other hand, the statistical significance to exclude wrong-octant hypothesis of \thetamu\ mixing angle from $\bar{\nu_{e}}$ \textit{appearance} sub-sample is higher in the parameter space ($\delta_{CP} \approx -\pi/2, \sin^{2}\theta_{23} < 0.45)$ and ($\delta_{CP} = \pi/2, \sin^{2}\theta_{23} > 0.55)$. These numerical results, along with a breakdown in sensitivity of T2HK, DUNE and ESSnuSB presented in Appendix~\ref{sec:AppnuAnti}, agree with expectation from analysis of oscillation probability of $\nu_{\mu}\rightarrow \nu_e$ and $\overline{\nu}_{\mu}\rightarrow \overline{\nu}_e$ transition presented in Appendix~\ref{sec:AppDeg}. As argued in Section~\ref{sec:samples}, a data analysis on both neutrino and anti-neutrino sub-samples will mitigate the influence of \dcp\ actual values, as well as neutrino mass ordering,  on the \thetamu\ precision measurement. This argument is substantiated by a nearly constant fraction of $\sin^2\theta_{23}$ to exclude wrong-octant at 3$\sigma$ C.L. as a function of true \dcp\ values, as presented in Fig.~\ref{fig:ORvsdcp}. These scenarios are examined, including (i) mass ordering is \emph{normal} and unknown, (ii) mass ordering is \emph{normal} and known, and (iii) mass ordering is \emph{inverted} and known. The results across these scenarios are comparable. 

\begin{figure*}
    \centering
    \includegraphics[width=0.485\linewidth]{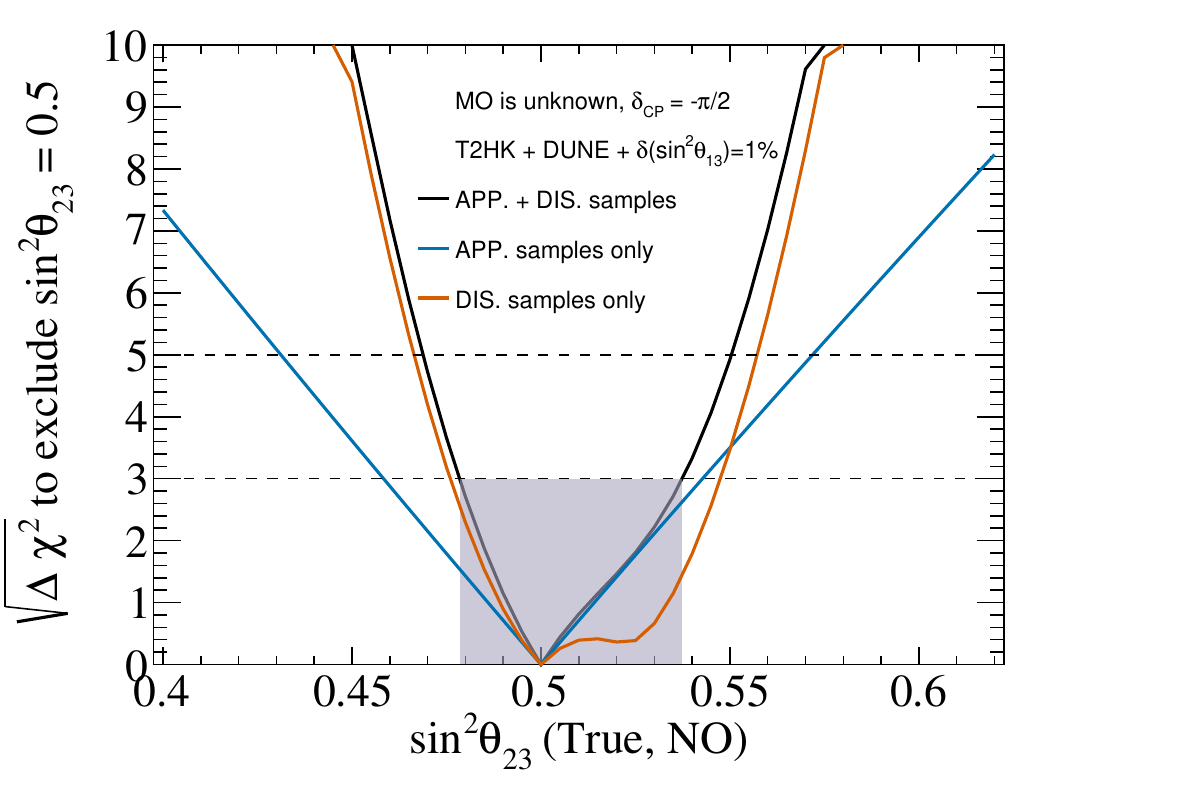}
    \includegraphics[width=0.485\linewidth]{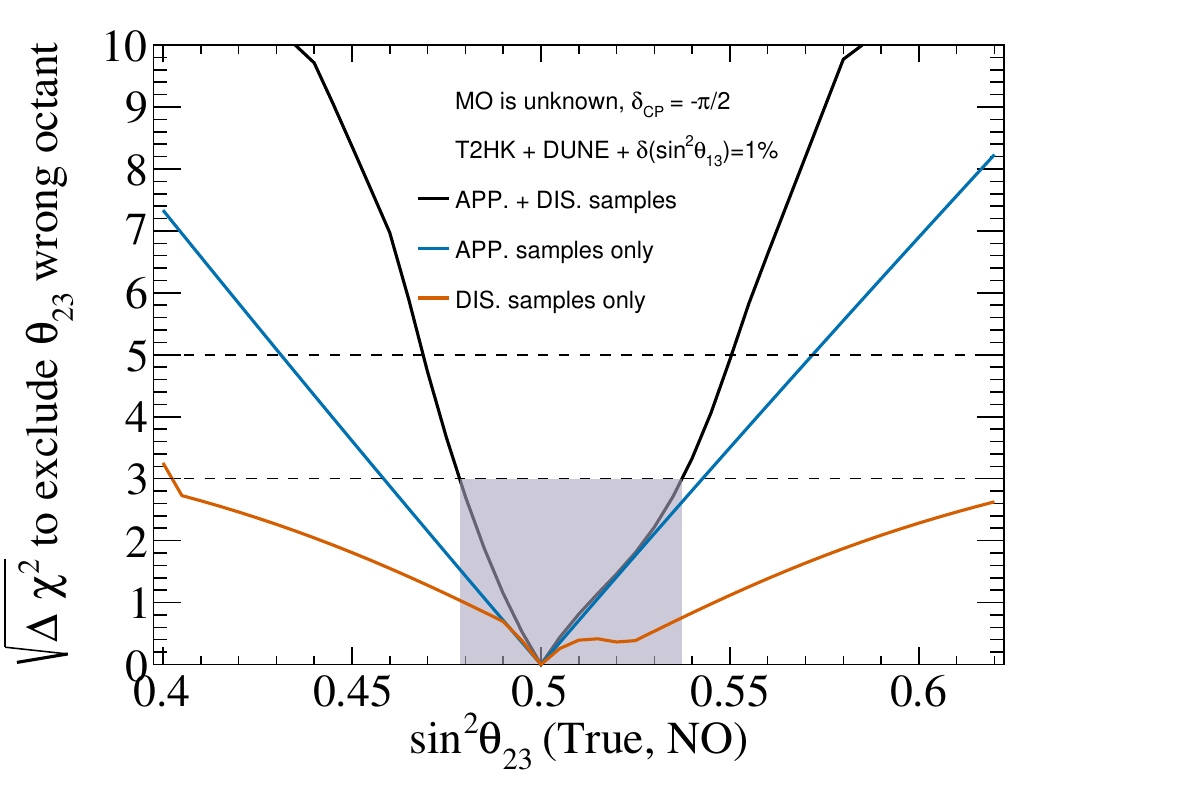}
    \caption{\label{fig:t2hkdune}Sensitivities to exclusion of the maximal-mixing hypothesis (left) and wrong-octant (right) are analyzed by separating or combining contribution from the \emph{appearance} and \emph{disappearance} sub-samples of T2HK and DUNE.  }
\end{figure*}

\noindent A potential joint analysis of T2HK and DUNE for separated \emph{appearance} and \emph{disappearance} sub-samples, as well as combined data, is conducted,  with results highlighted in Fig.~\ref{fig:t2hkdune}. In this context, sensitivities are calculated utilizing the ultimate 1\% uncertainty on $\sin^{2}\theta_{13}$~\citep{Zhang:2022zoc}, which can enhance the $3\sigma$ C.L. octant-blind regions constrained exclusively by T2HK and DUNE's \emph{appearance} samples by up to 11\%. For ruling out maximal-mixing, T2HK and DUNE with \emph{appearance} (\emph{disappearance}) sub-samples only can exclude the maximal-mixing with 3$\sigma$ C.L. for a fraction of 64\% (68\%) of the currently allowed range of the \thetamu\ angle, respectively. Notably, the \emph{appearance} sub-samples exhibit greater sensitivity if actual \thetamu\ lies in higher octant, whereas \emph{disappearance} sub-samples provide a superior contribution if actual \thetamu\ lies in lower octant. A combined \emph{appearance} and \emph{disappearance} datasets of T2HK and DUNE slightly enlarge the octant-resolving region of \thetamu\ up to 73\%. For excluding the wrong-octant, contribution of the \emph{appearance} sub-samples is superior over the \emph{disappearance} one for all possible values of the \thetamu\ angle. Specifically, T2HK and DUNE with \emph{appearance} sub-samples only are anticipated to conclude the octant of \thetamu\ angle if its actual value lies outside of the interval [0.46, 0.54], covering about 64\% of the currently allowed range of this parameter. A combination with both the \emph{disappearance} samples and \emph{appearance} sampples gain significantly if the actual value of \thetamu\ is in the lower octant, enlarging the octant-resolving region up to 73\% of the currently allowed region of \thetamu.  

\section{Conclusion}\label{sec:cons}
Precise measurement of the leptonic mixing angle \thetamu, alongside the investigation of CP violation and the determination of neutrino mass ordering, is considered a significant challenge in neutrino physics. Such efforts aim to complete the leptonic mixing framework and examine various flavor models. This study separately analyzes the contributions of $\overset{\brabar}{\nu}_{\mu}\rightarrow \overset{\brabar}{\nu}_{e}$ \emph{appearance} and $\overset{\brabar}{\nu}_{\mu}\rightarrow \overset{\brabar}{\nu}_{\mu}$ \emph{disappearance} data samples through oscillation probability and numerical simulation analyses. A combined data analysis can enhance parameter constraints; however, caution is necessary regarding the underlying assumptions for a unified data description. Our findings indicate that, the \emph{appearance} samples are the primary contributor to octant resolution across the entire range of allowed parameter values. Additionally, it significantly excludes the maximal-mixing hypothesis if the true value of the \thetamu\ mixing angle is situated in the higher octant and near $\pi/4$. The \emph{disappearance} samples are crucial for excluding the maximal-mixing hypothesis if the actual value is located in the lower octant or in the higher octant but distant from $\pi/4$. For broad-band neutrino experiments like DUNE, the contribution of \emph{disappearance} samples to the octant sensitivity remains substantial when \thetamu\ is close to the currently-allowed 3$\sigma$ bound. Utilizing only the T2HK and DUNE's \emph{appearance} samples allows for the determination of the octant of the \thetamu\ mixing angle for approximately 62\% of the currently allowed range of this parameter. We present an argument based on the analytical formula for oscillation probability and validate it through numerical simulations of the neutrino experiment, revealing that the precise measurement of the \thetamu\ mixing angle remains unaffected by the actual value of the CP violation phase and the unknown neutrino mass ordering. 

\section*{Acknowledgements}
S. Cao would like to thank Neutrino Group, IPNS, KEK for their hospitality and support during his visit. Phan To Quyen was funded by the Master, PhD Scholarship Programme of Vingroup Innovation Foundation (VINIF), code VINIF.2023.TS.095. The research of S. Cao and N. T. H. Van is funded by the National Foundation for Science and Technology Development (NAFOSTED) of Vietnam under Grant No. 103.99-2023.144. 

\bibliographystyle{unsrt}
\bibliography{Reference}
%
 \appendix
\section{Parameter degeneracy in the \emph{appearance} probabilities}\label{sec:AppDeg}
\begin{figure*}
\includegraphics[width=0.495\textwidth]{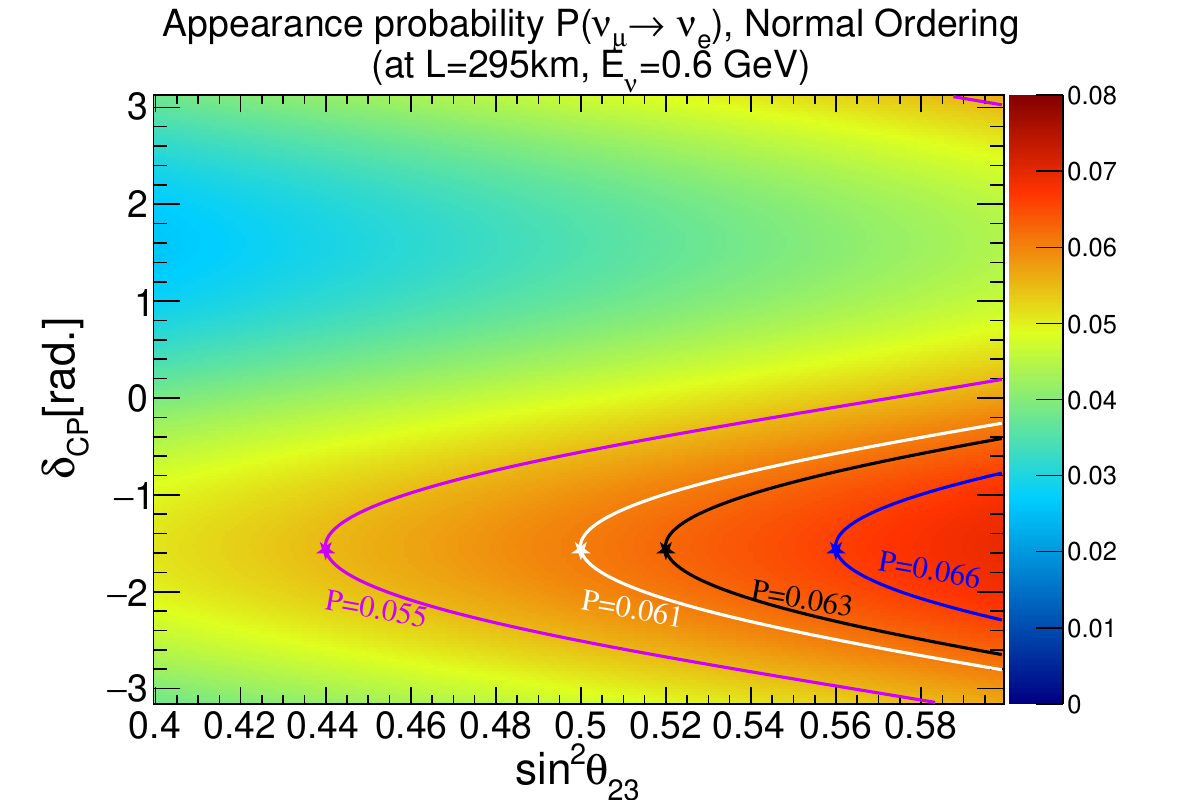}
\includegraphics[width=0.495\textwidth]{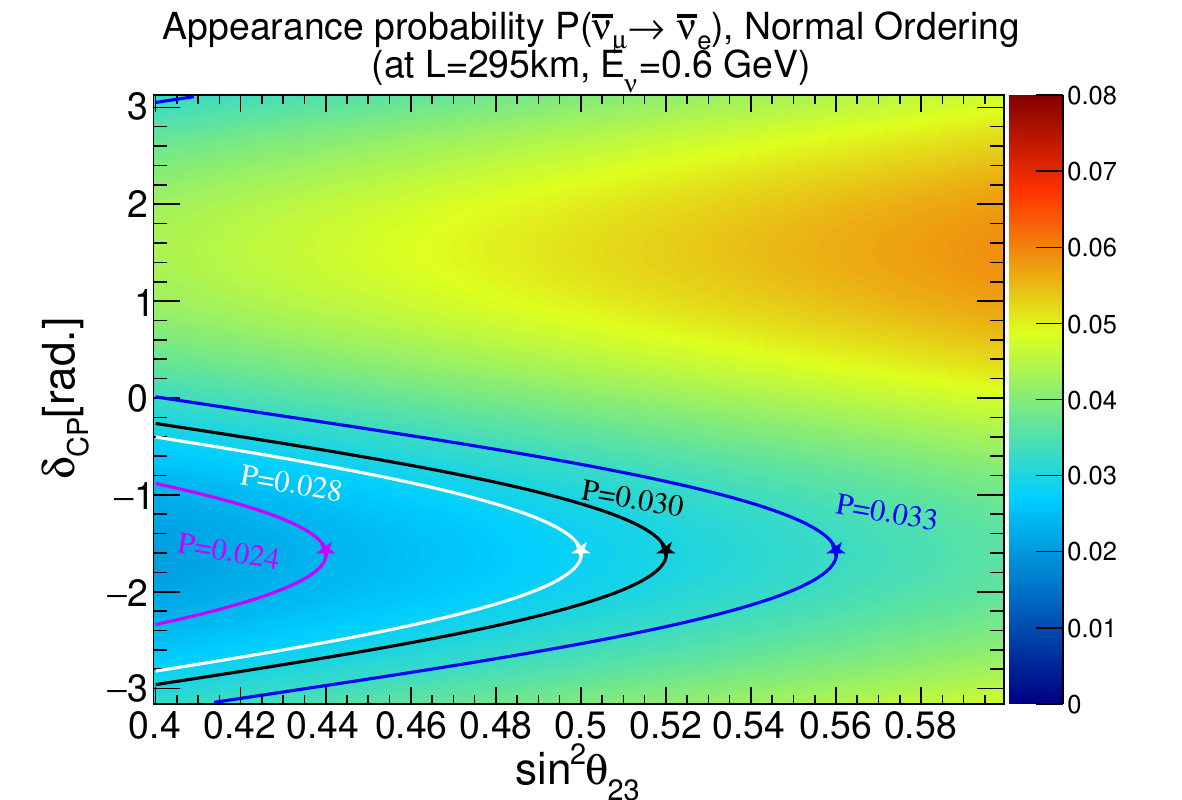}
\includegraphics[width=0.495\textwidth]{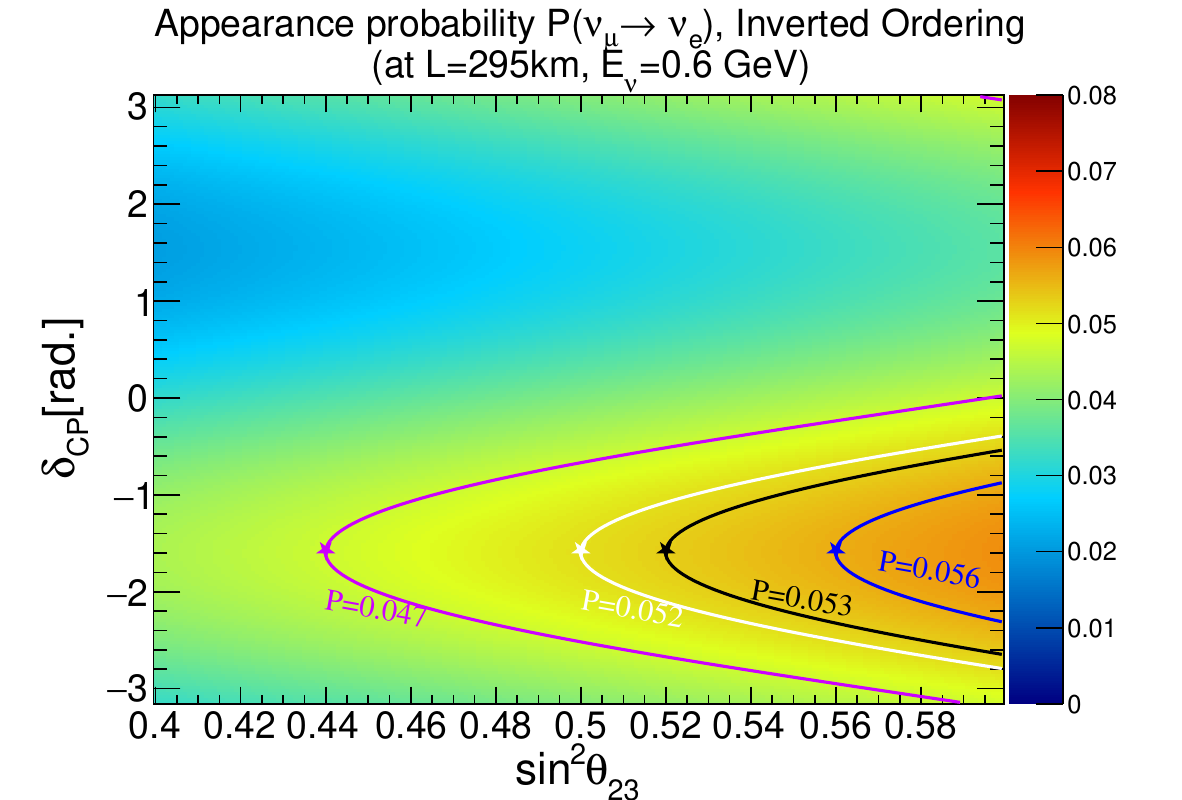}
\includegraphics[width=0.495\textwidth]{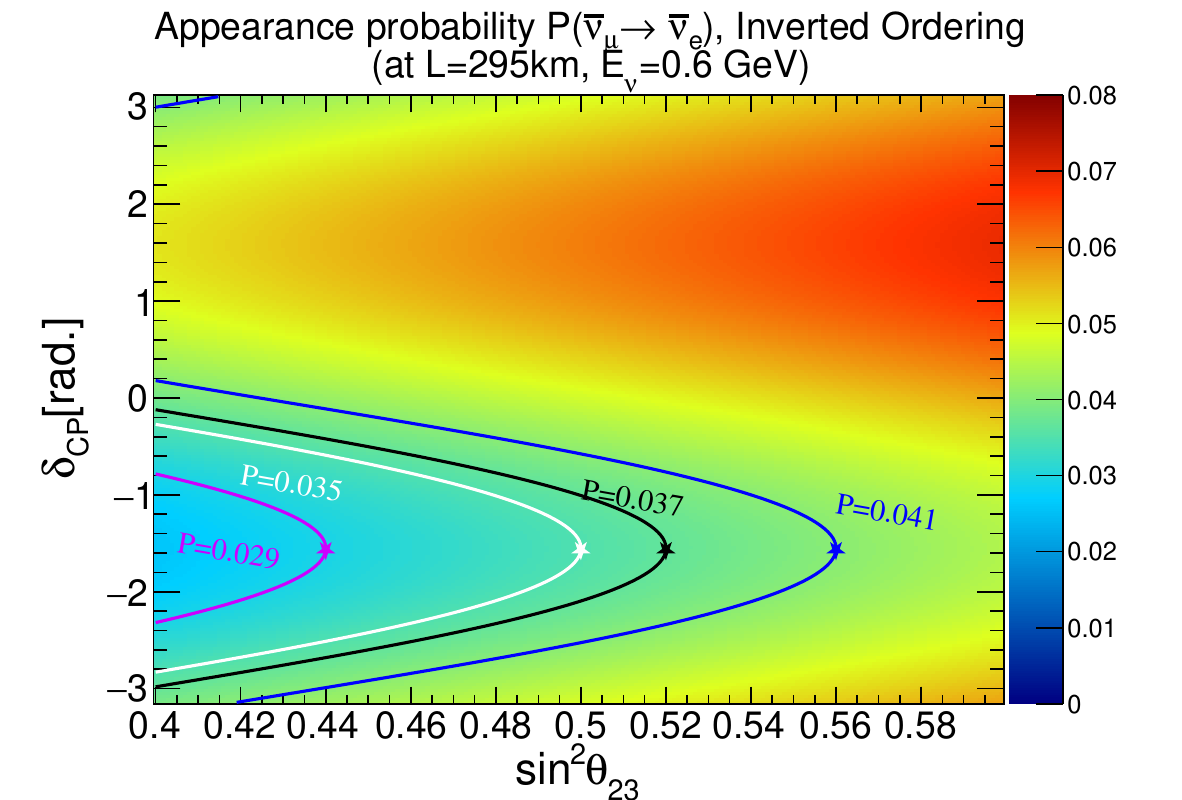}
\caption{\label{fig:prob_numu2nue_th23vsdcp} The  $\nu_{e}$ and $\overline{\nu}_e$ \emph{appearance} probabilities are presented on the $\sin^{2}\theta_{23}-\delta_{CP}$ parameter space. The iso-probability curves are calculated at four points of $\sin^2\theta_{23}$=(0.44, 0.50, 0.52, 0.56) and $\delta_{CP} = -\pi/2$. Other oscillation parameters are fixed as NuFit 5.2.}
\end{figure*}

\begin{figure*}
\includegraphics[width=0.495\textwidth]{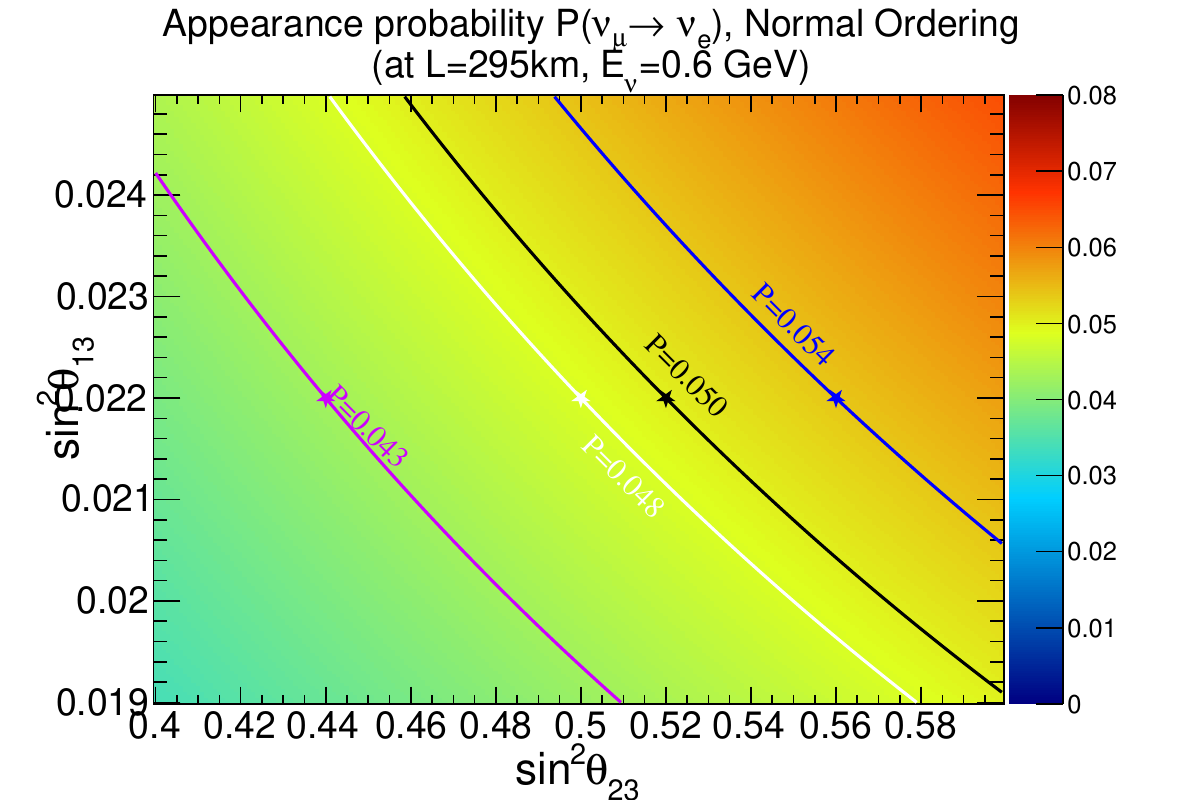}
\includegraphics[width=0.495\textwidth]{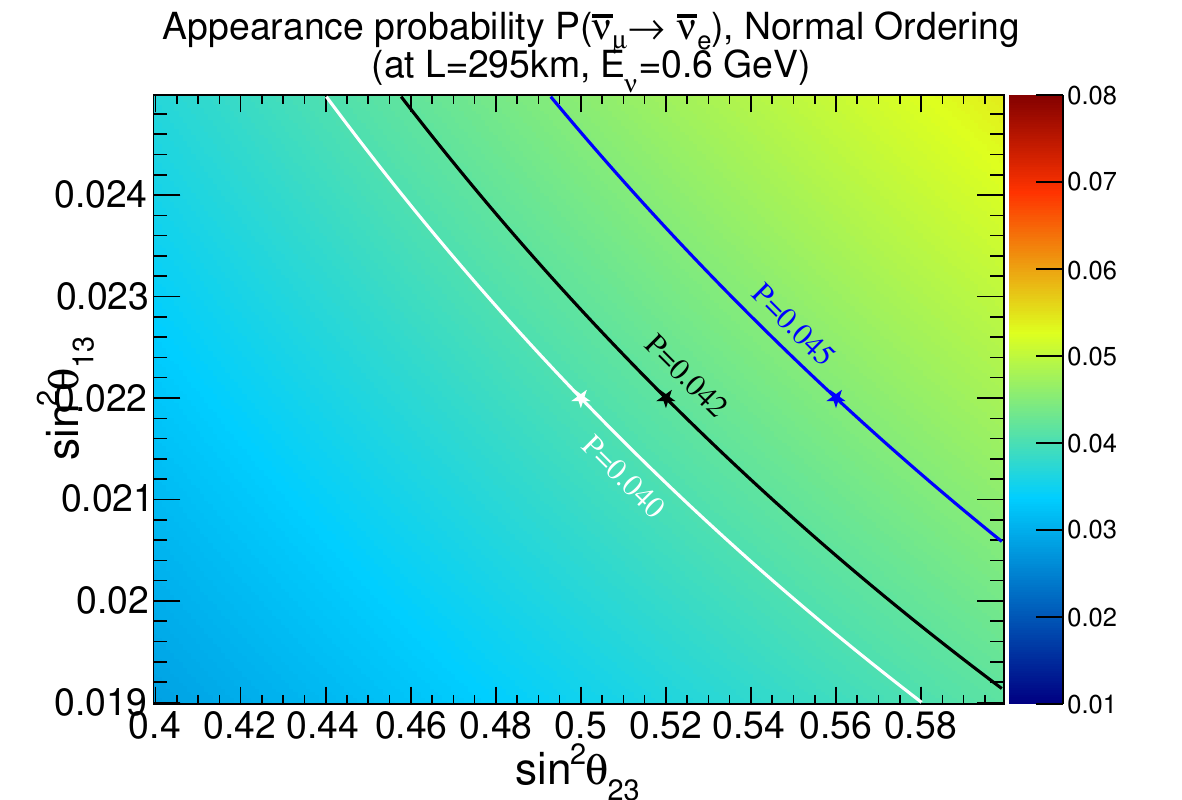}
\caption{\label{fig:prob_numu2nue_th23vsth13} The  $\nu_{e}$ and $\overline{\nu}_{e}$ \emph{appearance} probabilities are presented on the $\sin^{2}\theta_{23}-\sin^{2}\theta_{13}$ parameter space. The iso-probability curves are calculated at four points of $\sin^2\theta_{23}$=(0.44, 0.50, 0.52, 0.56). Other oscillation parameters are fixed as NuFit 5.2.}
\end{figure*}

The $\nu_{e}$ ($\overline{\nu}_e$) \emph{appearance} probabilities as function of ($\sin^{2}\theta_{23}-\delta_{CP}$) are shown in Fig.~\ref{fig:prob_numu2nue_th23vsdcp}. The degeneracy in ($\sin^{2}\theta_{23}-\delta_{CP}$) solution sounds pronounced when looking at the iso-probability curves for each individual channel. However, one can notice that the intersection of the iso-probability curves for the neutrino (the left of Fig.~\ref{fig:prob_numu2nue_th23vsdcp}) and anti-neutrino (the right of Fig.~\ref{fig:prob_numu2nue_th23vsdcp}) are right at the truth solutions. It indicates that $\nu$ and $\bar{\nu}$ appearance samples share a mirror symmetry and then cancel out if combine neutrino and anti-neutrino. Therefore, it shows the independence of the truth $\delta_{CP}$ value in the precise measurement of $\sin^{2}\theta_{23}$ which we observe in Fig.\ref{fig:prob_numu2numu}. Moreover, it is found that the relative difference between the $\nu_{e}$ ($\overline{\nu}_{e}$ \emph{appearance}) probabilities in the \emph{normal} and \emph{inverted} mass ordering is almost independent on the true value of $\sin^{2}\theta_{23}$ but only the true value of $\delta_{CP}$. Thus it leads to a consequence that precision measurement of $\sin^{2}\theta_{23}$ does not depend on the neutrino mass ordering, which is deeply connected to the degeneracy decoupling of these two parameters in the $\nu_{\mu}\rightarrow \nu_e$ oscillation measurement~\citep{Kajita:2006bt}.
\noindent Fig.\ref{fig:prob_numu2nue_th23vsth13} shows $\nu_{e} (\overline{\nu}_{e})$ appearance probabilities as a function of $(\sin^{2}\theta_{23} - \sin^{2}\theta_{13}$). The degeneracy between $\theta_{13} - \theta_{23}$ are taken from Eq.\ref{eq:probnumutonue} and given as Eq in page 3 \citep{Cao:2024ptn}. It shows that the effect of the $\theta_{23}$ octant determination in case of non-maximal mixing is significant
from the precision value of $\theta_{13}$. 

\section{\label{sec:AppnuAnti}Contribution of neutrino and anti-neutrino data samples}
\begin{figure*}
\includegraphics[width=0.495\textwidth]{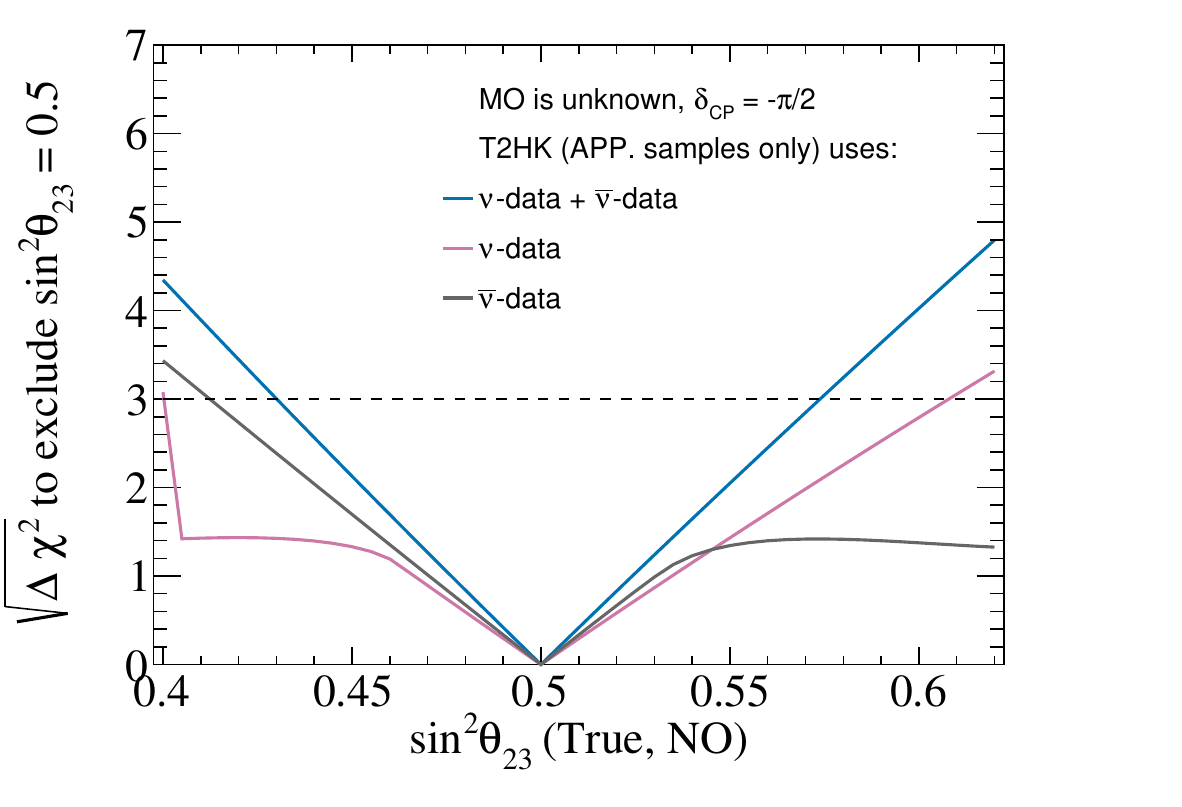}
\includegraphics[width=0.495\textwidth]{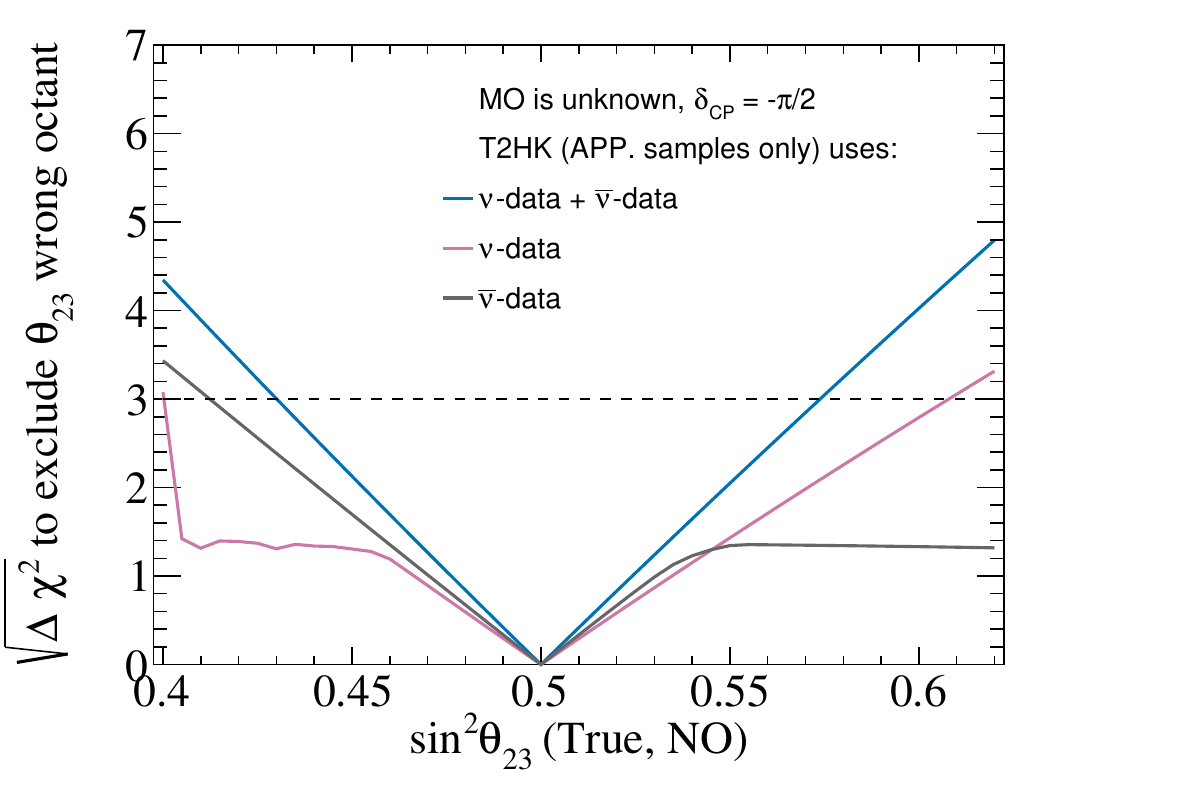}
\includegraphics[width=0.495\textwidth]{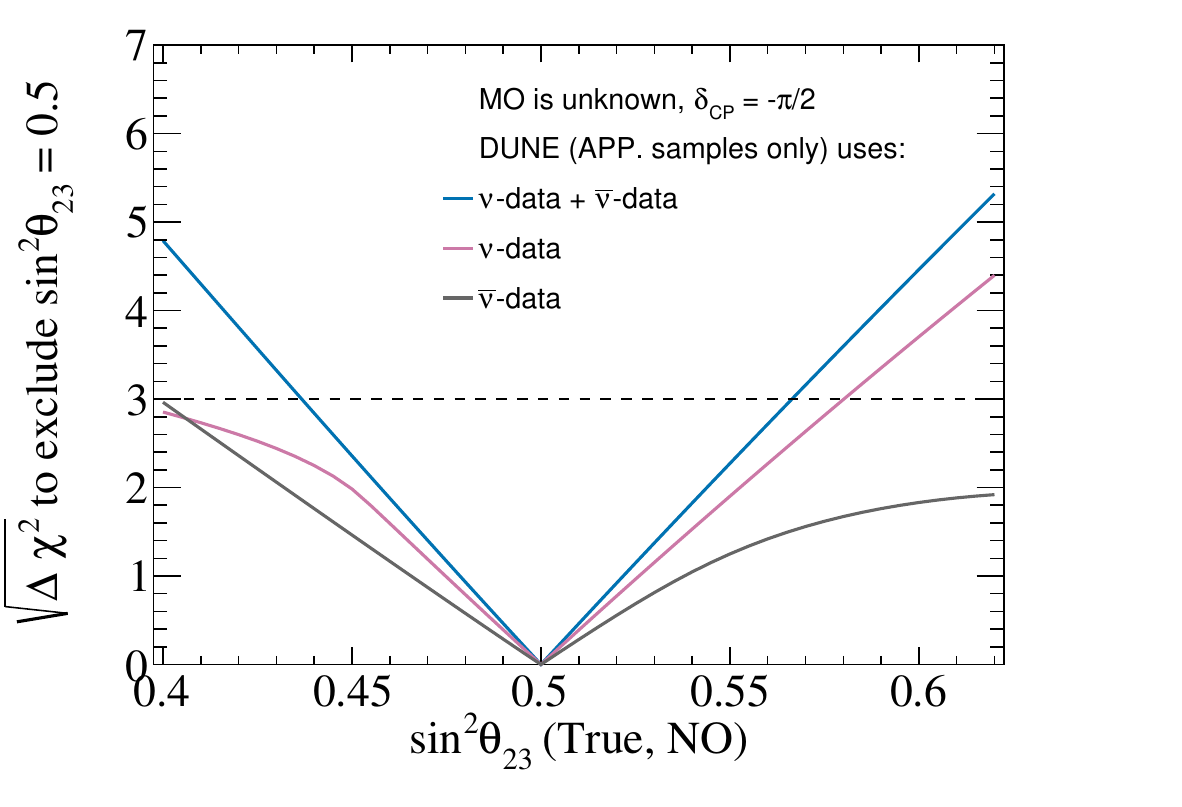}
\includegraphics[width=0.495\textwidth]{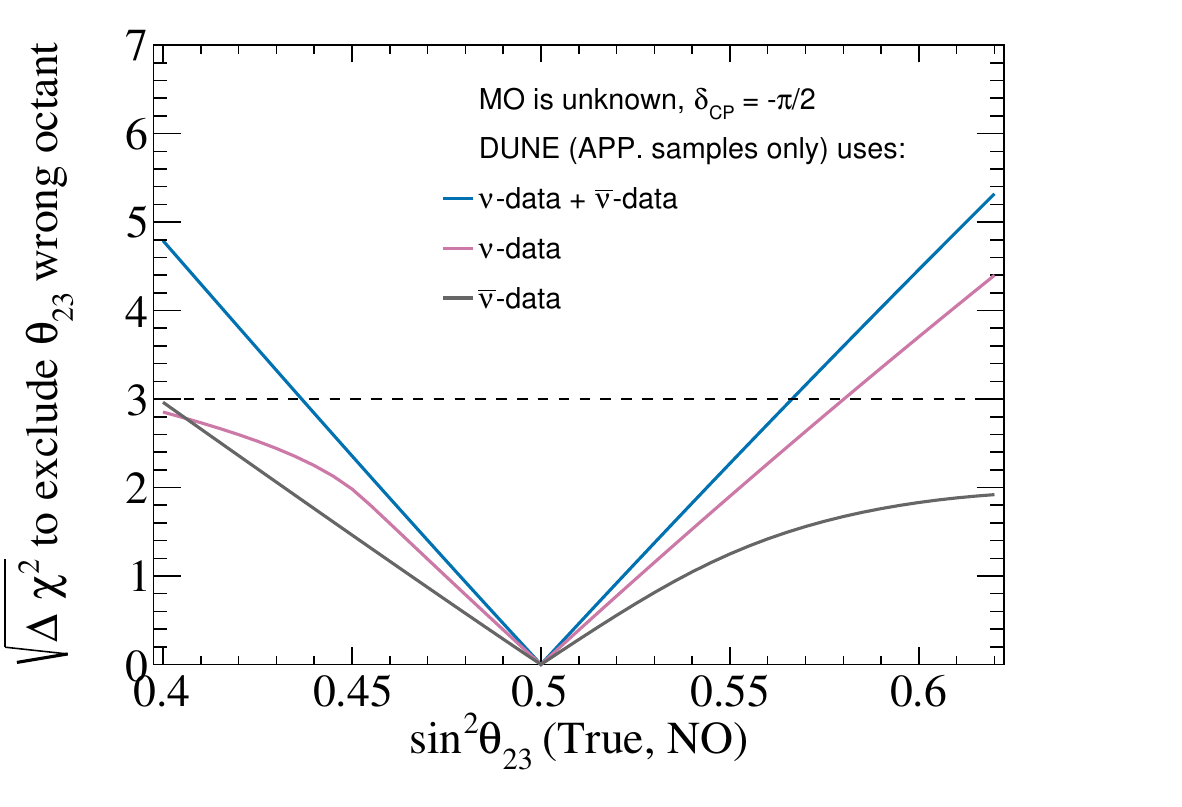}
\includegraphics[width=0.495\textwidth]{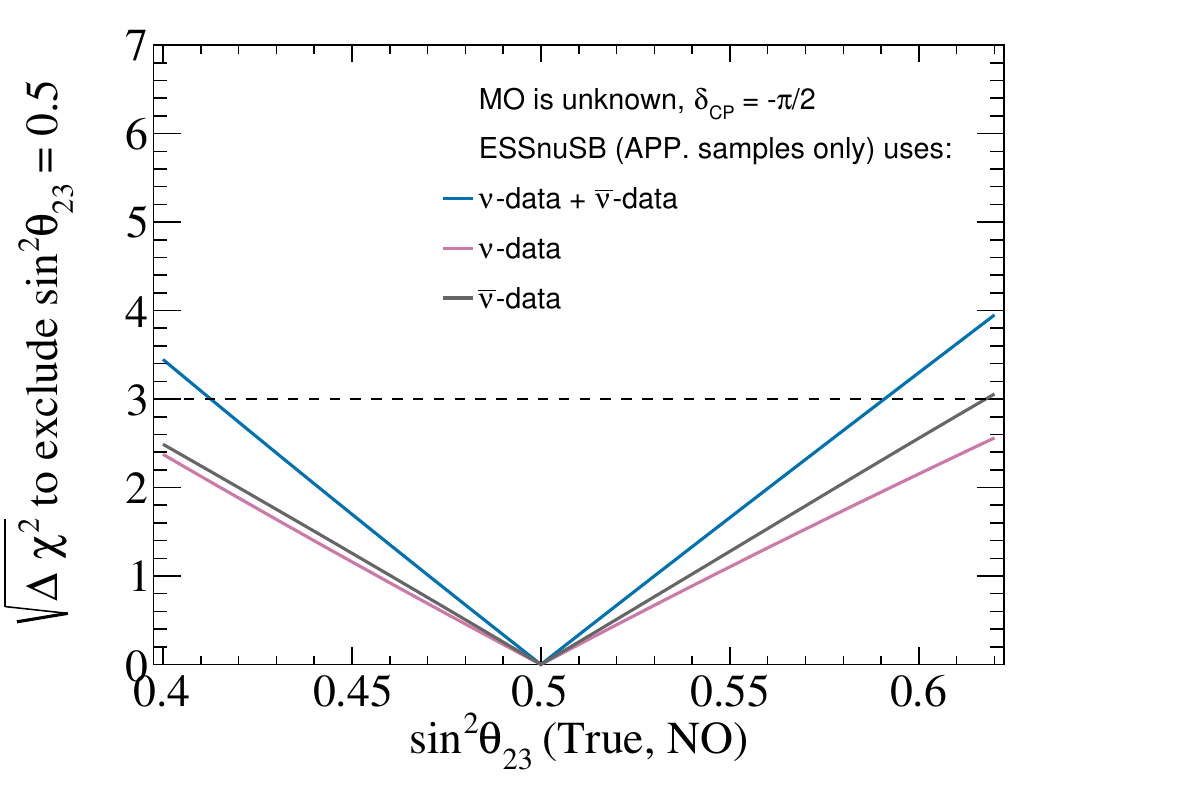}
\includegraphics[width=0.495\textwidth]{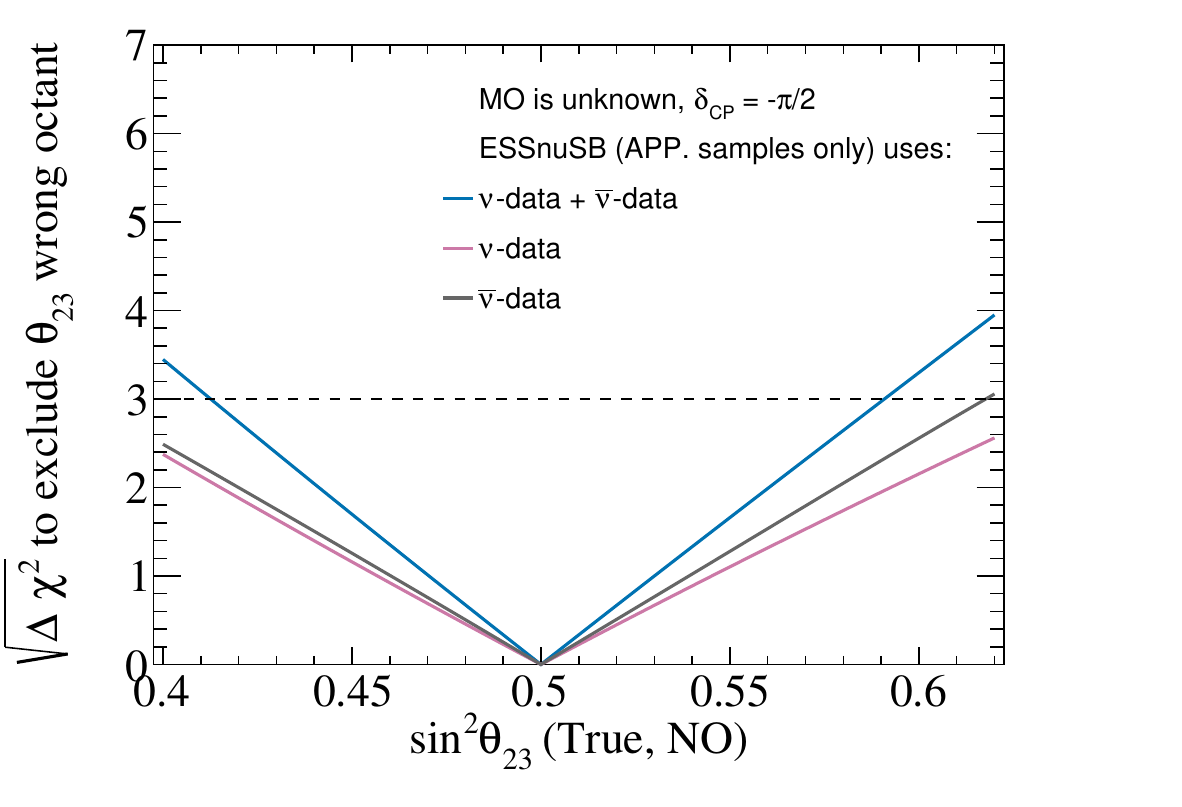}
\caption{\label{fig:twotestnuantinu} The $\nu$ and $\bar{\nu}$ statistical significance to exclude the maximal mixing (left) and wrong-octant (right) as function of $\sin^{2}\theta_{23}$ using the appearance samples only. Here $\delta_{CP} = -\pi/2$ is set, MO is presumably unknown, and other relevant parameters and their uncertainties are taken from Table~\ref{tab:nuoscpara}.} 
\end{figure*}

We investigate the separated contributions of $\nu$ and $\bar{\nu}$ sub-samples in \emph{appearance} datasets for determining the true octant and excluding the maximal mixing hypothesis. The results, presented in Fig.~\ref{fig:twotestnuantinu}, show that $\nu_{e}$ and $\overline{\nu}_{e}$ \emph{appearance} samples provide different sensitivities in measuring $\theta_{23}$, but the combination of samples is useful to enhance the statistics collected from experiments, leading to more precise $\theta_{23}$ measurements.

\end{document}